\documentclass[a4paper,12pt]{article}

\usepackage{putex}

\usepackage[colorlinks=true,urlcolor=blue,anchorcolor=blue,citecolor=blue,filecolor=blue,linkcolor=blue,menucolor=blue,pagecolor=blue,linktocpage=true]{hyperref}
\usepackage{mathtools}
\usepackage{amsfonts}
\usepackage{amssymb}
\usepackage{bbm}
\usepackage{amsthm}
\usepackage{cite}
\usepackage{MnSymbol}

\newcommand{\Of}{O_f}
\newcommand{\Ob}{O_b}
\newcommand{\Jf}{J^f}
\newcommand{\Jb}{J^b}
\newcommand{\vphi}{\varphi}
\newcommand{\ed}{\,.}
\newcommand{\ec}{\,,}
\newcommand{\ecq}{\ec\quad}

\newcommand{\dho}{\partial}
\newcommand{\tl}{\tilde{\lambda}}
\newcommand{\la}{\langle}
\newcommand{\ra}{\rangle}

\newcommand{\cBcrit}{\ensuremath{\mathcal{B}^{\mathrm{crit.}}}}
\newcommand{\cBreg}{\ensuremath{\mathcal{B}}}
\newcommand{\tOb}{\widetilde{O}_b}

\newcommand{\bZ}{\ensuremath{\mathbb{Z}}}

\newcommand{\cB}{\ensuremath{\mathcal{B}}}

\newcommand{\cD}{\ensuremath{\mathcal{D}}}

\newcommand{\cF}{\ensuremath{\mathcal{F}}}

\newcommand{\cJ}{\ensuremath{\mathcal{J}}}

\newcommand{\cM}{\ensuremath{\mathcal{M}}}
\newcommand{\cN}{\ensuremath{\mathcal{N}}}
\newcommand{\cO}{\ensuremath{\mathcal{O}}}

\newcommand{\cT}{\ensuremath{\mathcal{T}}}

\newcommand{\cV}{\ensuremath{\mathcal{V}}}

\begin{document}

\preprint{PUPT-2482}

\institution{Stanford}{
Stanford Institute for Theoretical Physics, 
Stanford University, Stanford, CA 94305, USA}

\institution{Princeton}{
Joseph Henry Laboratories, Princeton University, Princeton, NJ 08544, USA}

\title{Three Dimensional Bosonization From Supersymmetry}

\authors{Guy Gur-Ari,\worksat{\Stanford}\footnote{e-mail: {\tt guyga@stanford.edu}} and Ran Yacoby\worksat{\Princeton}\footnote{e-mail: {\tt ryacoby@princeton.edu}}}

\abstract{
Three dimensional bosonization is a conjectured duality between non-supersymmetric Chern-Simons theories coupled to matter fields in the fundamental representation of the gauge group.
There is a well-established supersymmetric version of this duality, which involves Chern-Simons theories with $\cN=2$ supersymmetry coupled to fundamental chiral multiplets.
Assuming that the supersymmetric duality is valid, we prove that non-supersymmetric bosonization holds for all planar correlators of single-trace operators.
The main tool we employ is a double-trace flow from the supersymmetric theory to an IR fixed point, in which the scalars and fermions are effectively decoupled in the planar limit.
A generalization of this technique can be used to derive the duality mapping of all renormalizable couplings, in non-supersymmetric theories with both a scalar and a fermion.
Our results do not rely on an explicit computation of planar diagrams.
}

\maketitle

\tableofcontents

\setlength{\unitlength}{1mm}

\section{Introduction and Summary of Results}
\label{intro}

Bosonization in three-dimensional quantum field theories is a recently conjectured duality, first developed in \cite{Giombi:2011kc,Aharony:2011jz,Maldacena:2011jn,Maldacena:2012sf,Aharony:2012nh}, between certain Chern-Simons theories with $U(N)$ or $O(N)$ gauge groups, coupled to matter fields in the fundamental representation.\footnote{It should be straightforward to extend these dualities to $USp(2N)$ gauge theories as well.} We will refer to this class of theories as Chern-Simons vector models. The most basic example of 3d bosonization states that a Chern-Simons theory coupled to a single fermion $\psi(x)$ (the `fermionic' model) is equivalent to another such theory coupled to a scalar $\vphi(x)$ at the Wilson-Fisher fixed point (the `critical bosonic' model). This was later generalized in \cite{Aharony:2012ns,Jain:2013gza} to a duality of a Chern-Simons vector model containing both a fermion and a scalar, with the most general renormalizable matter potential $V(\vphi,\psi)$.\footnote{In this case the dual theory is also a Chern-Simons vector model with a fermion and a scalar. The justification for the name `bosonization' in this theory is that  the duality maps operators constructed from fermion bilinears to those constructed from scalars.} For a particular choice of $V(\vphi,\psi)$ the theory has $\cN=2$ supersymmetry, and bosonization becomes equivalent to a Seiberg-like duality, which was first proposed for a general class of $\cN=2$ Chern-Simons vector models by Giveon and Kutasov \cite{Giveon:2008zn} and later generalized to the particular case described above in \cite{Benini:2011mf} (see also \cite{Kapustin:2011gh,Willett:2011gp}).

So far, the only direct evidence supporting bosonization comes from computations done in the large $N$ limit \cite{Giombi:2011kc,Aharony:2011jz,Maldacena:2011jn,Maldacena:2012sf,Aharony:2012nh,GurAri:2012is,Aharony:2012ns,Jain:2012qi,Jain:2013gza,Jain:2013py,Takimi:2013zca,Yokoyama:2013pxa,Jain:2014nza,Gurucharan:2014cva,Dandekar:2014era,Moshe:2014bja,Bedhotiya:2015uga,Inbasekar:2015tsa} (see also \cite{Yokoyama:2012fa,Banerjee:2012gh,Frishman:2013dvg,Banerjee:2013nca,Bardeen:2014paa,Bardeen:2014qua,Frishman:2014cma} for related work). On the other hand, Giveon-Kutasov duality was tested at finite $N$. Because the $\cN=2$ duality is much more well established than the non-supersymmetric one, studying the relation between the two may give rise to new evidence for 3d bosonization. Motivated by this possibility, in this work we will explore the relation between the two dualities. In particular, the main question we wish to address is whether the non-supersymmetric bosonization dualities described above can be derived from the Giveon-Kutasov duality of the $\cN=2$ $U(N)$ Chern-Simons theory with a single fundamental chiral multiplet. 

 The idea that 3d bosonization can be derived from Giveon-Kutasov duality was already exhibited in the important paper \cite{Jain:2013gza}. In that work, the planar thermal free energy of the $\cN=2$ theory was computed in the presence of relevant, supersymmetry-breaking deformations. Sending the coupling of one such deformation to infinity decouples the fermion and leads to the planar free energy of the critical bosonic model; the corresponding limit in the dual theory decouples the scalar, leading to the free energy of the fermionic model. The results of \cite{Jain:2013gza} constitute strong evidence that the non-supersymmetric theories are related to the $\cN=2$ theory by dual RG flows, at least when $N$ is sufficiently large. 

In the present work we will follow a different route. We will derive the non-supersymmetric duality from the supersymmetric one for a large class of observables, without computing these observables explicitly. Our main results can be stated as follows. If the duality of the $\cN = 2$ theory is valid, then all correlation functions of single-trace operators in the critical bosonic and fermionic models agree under the bosonization duality in the planar limit.  
We will also give a simple derivation of the large $N$ duality map of the coupling constants in the most general renormalizable potential $V(\vphi,\psi)$, in the theory that contains both a scalar and a fermion. The duality map for some of those couplings was not known previously. 

Note that in the critical bosonic and fermionic models, all planar 2-point and 3-point functions of single-trace operators are known explicitly, and were already shown to match under bosonization. 
Our results also apply to higher-point functions, whose form is not generally known.\footnote{See \cite{Bedhotiya:2015uga} for an exception.} We stress that planar, connected $n$-point functions of single-trace operators are generally independent observables for any $n$, even in conformal field theories (CFTs).\footnote{Indeed, upon using the operator product expansion, such correlators with $n\geq 4$ contain non-trivial contributions from 3-point functions with multi-trace operators; these were never computed explicitly in the theories discussed in this work.} Therefore, one can view our result as new evidence for bosonization, which goes beyond the one provided by matching the known planar 2-point and 3-point functions.

Let us now describe our derivation of the duality between the critical bosonic and fermionic models, which we will approach in two different ways. The first method uses a perturbative expansion in the multi-trace interactions of the $\cN=2$ action. (Note that this is not ordinary perturbation theory, because we treat the gauge interactions exactly.) We begin by writing down the duality equation for certain correlators of the $\cN=2$ theory, and expanding them perturbatively. By re-arranging the perturbative contributions we obtain an equality between correlators of the critical bosonic and fermionic theories, in agreement with the bosonization duality.\footnote{This argument is close in spirit to an argument made in \cite{Aharony:2011jz}, where it was shown that certain scalar operators in the bosonic and fermionic theories do not acquire an anomalous dimension. This was done by perturbatively relating the 2-point functions of these operators to similar 2-point functions in the $\cN=2$ theory, where supersymmetry implies that the corresponding anomalous dimensions vanish.}
In the large $N$ limit the perturbative expansion converges and the result is exact. We will use this method to derive the bosonization duality of the 4-point function of spin 1 currents, as well as of all 3-point functions (a known result). We believe that it is possible to extend the same argument to other higher-point functions, but this becomes tedious. Instead, in Section \ref{dualityProof} we give a simpler derivation of 3d bosonization that we will discuss in the rest of the introduction.

Our $\cN=2$ theory contains a scalar field $\varphi$ and a Dirac fermion $\psi$, both in the fundamental of the $U(N)$ gauge group. 
The gauge-invariant operator $\bar{\vphi}\vphi$ has vanishing anomalous dimension, because it sits in the same multiplet as the conserved $U(1)$ flavor current.
Therefore, the double-trace operator $(\bar{\varphi} \varphi)^2$ is relevant for large enough $N$.
In the planar limit we may deform the supersymmetric theory by $(\bar{\varphi} \varphi)^2$ and flow to an IR fixed point that is not supersymmetric.
The IR theory includes a fermion and a Wilson-Fisher scalar, both coupled to a gauge field with Chern-Simons interactions.\footnote{The duality of this non-supersymmetric theory was already considered at the level of the thermal free energy in \cite{Jain:2013gza}.}
The Giveon-Kutasov duality of the UV theory becomes a duality of the non-supersymmetric IR theory.

As we will show, in the planar limit of the IR theory the scalar and fermion are effectively decoupled for a large class of observables.
In particular, planar correlators of single-trace operators that are composed of the scalar $\varphi$ and the gauge field do not receive contributions from interactions with the fermion $\psi$. These correlators are therefore equal to those of the critical bosonic theory. Similarly, correlators of single-trace operators that involve only fermions and gauge fields are equal to those of the fermionic theory. Using this decoupling we will derive the duality of the critical bosonic and fermionic theories. For this derivation to work, we must know the mapping of all single-trace operators under the Giveon-Kutasov duality. We will determine this map by working out the arrangement of these operators inside multiplets of the $\cN=2$ superconformal algebra. In the process, we will uncover signs in the duality map that were not noted previously.

The way in which the double-trace deformation $\int \! d^3x \, g (\bar{\varphi} \varphi)^2$ is embedded within a supersymmetric deformation plays an important role in our argument. To understand this embedding, we will use the fact that the IR CFT at the end of the double-trace flow can be equivalently described by coupling the operator $\bar{\varphi} \varphi$ to a background field $\hat{D}$ in the UV theory, and then making $\hat{D}$ dynamical; this equivalence can be seen by using the Hubbard-Stratonovich trick. In the $\cN=2$ theory, the background field $\hat{D}$ is the top component of the background vector multiplet that contains the $U(1)$ flavor current.
It follows that the supersymmetric completion of making a double-trace deformation and flowing to the IR CFT involves gauging the flavor $U(1)$ symmetry.\footnote{The Hubbard-Stratonovich transformations also contains the deformation $\int \! d^3x \frac{1}{4 g} \hat{D}^2$ that can be supersymmetrized to an $\cN=2$ Yang-Mills term. This deformation is irrelevant and can be ignored in the IR CFT.} Because we know how the flavor symmetry maps under Giveon-Kutasov duality, we can work out the exact mapping of our supersymmetry-breaking deformation. 

We see that supersymmetry allows us to map the double-trace deformation across the duality.
It is possible to extend this basic strategy to additional deformations by making other fields in the background vector multiplet dynamical with some particular weights, analogous to the $\hat{D}^2$ term in the Hubbard-Stratonovich transformation. In particular, we will use this strategy to derive the large $N$ duality map of the $U(N)$ Chern-Simons vector model containing both a scalar and a fermion with the most general renormalizable potential $V(\vphi,\psi)$. In \cite{Aharony:2012ns,Jain:2013gza}, the duality map of those couplings in $V(\vphi,\psi)$ that contribute to the planar thermal free energy was determined. We find perfect agreement with \cite{Jain:2013gza}, and also extend their results to all the other couplings in $V(\vphi,\psi)$. The advantage of our method is that it provides a very simple derivation of the duality map, which does not require performing any complicated all-order computations.

There is one important subtlety in making background fields dynamical. Local terms in the action that are non-linear in the background fields contribute to contact terms of the correlation functions generated by those fields. Once the background fields are made dynamical, such terms become ordinary kinetic or interaction terms that can affect the correlators of the new theory even at separated points. The upshot is that, in order to derive a new duality using this strategy, one must make sure that the duality in the original theory extends to certain contact terms. In the present context, we will see that a crucial role in our derivation is played by the global Chern-Simons term of the background vector multiplet corresponding to the flavor $U(1)$ symmetry, which must be added for the validity of the $\cN=2$ duality \cite{Benini:2011mf,Closset:2012vp}.

The paper is structured as follows. In Section \ref{setup} we present the Chern-Simons vector models and explain how planar 2-point correlators in the $\cN=2$ theory are related to those of the non-supersymmetric theories in perturbation theory. In Section \ref{N2duality} we determine the $\cN=2$ duality map of all the bosonic single-trace operators. In Sections \ref{cpt} and \ref{dualityProof} we then prove the non-supersymmetric duality for planar correlators in two different ways, using perturbation theory in the $\cN=2$ multi-trace couplings, and using a double-trace flow. In Section \ref{BosFer} we derive the mapping of all renormalizable couplings under the Giveon-Kutasov duality. Section \ref{discussion} contains a discussion of our results, and the Appendices contain our conventions and some technical proofs.

\section{Preliminaries}
\label{setup}

In this section we will define the field theories discussed in this work, give a review of their single-trace spectrum, and provide some properties of 2-point functions that will be needed in later sections. More details on our conventions can be found in Appendix \ref{conv}.

\subsection{$U(N)$ Chern-Simons Vector Models}

Let us define the three main field theories to be discussed in this work. These are all Chern-Simons theories with $U(N)$ gauge group at level $k$, coupled to matter in the fundamental representation of $U(N)$.\footnote{For a Yang-Mills-Chern-Simons theory with gauge group $G$ and bare level $k_0$, the renormalized Chern-Simons level in the IR is given by $|k|=|k_0| + h^{\vee}$, where $h^{\vee}$ is the dual coxeter number of $G$. In this work we exclusively use $k$ to denote this renormalized level. In particular, for the $U(N)$ gauge group $h^{\vee}=N$, which implies that $|k| \ge N$.}  The different theories are distinguished by their matter sector: 
\begin{itemize}
	\item The theory $\cF_{k,N}$ has a single Dirac fermion $\psi(x)$, and is commonly referred to as the regular fermion theory. Its Euclidean flat-space action is given by\footnote{The gauge field is $A_{\mu}=A_{\mu}^a T^a$ where $T^a$ ($a=1,\ldots,N^2$) are hermitian generators of the gauge symmetry algebra in the fundamental representation, normalized such that $\tr_N(T^aT^b)=\frac{1}{2}\delta^{ab}$.  The gauge covariant derivative acts on fundamentals as $\cD_{\mu}\vphi=\dho_{\mu}\vphi-iA_{\mu}\vphi$, and on anti-fundamentals as $\cD_{\mu}\bar{\vphi} = \dho_{\mu}\bar{\vphi}+i\bar{\vphi}A_{\mu}$. More details on our conventions  are given in Appendix \ref{conv}.} 
	\begin{align}
	  S^{\cF}_{k,N}(A_{\mu},\psi) = \frac{ik}{4\pi}S_{\rm CS}(A_{\mu}) + S_f(A_{\mu},\psi) \ec \label{Sf}
	\end{align}
  where
  	\begin{align}
  	  S_{\rm CS}(A_{\mu}) &= 
  	  \int \! d^3x \, \epsilon^{\mu\nu\rho} \tr_N \left( 
  	  A_\mu \dho_\nu A_\rho - \frac{2i}{3} A_\mu A_\nu A_\rho
  	 \right) \ec \label{SCS}\\
  	  S_f(A_{\mu},\psi) &= -i\int \! d^3x \, \bar{\psi} \gamma^\mu \cD_\mu \psi
  	  \ed \label{Sff}
  	\end{align}
	
	\item The theory $\cBcrit_{k,N}$ has a single complex Wilson-Fisher scalar $\vphi(x)$, and is commonly referred to as the critical boson theory. One can flow to it by starting with the regular boson theory, deforming by a relevant double-trace interaction $\delta S = \int d^3x \frac{\tilde{\lambda}_4}{2N}(\bar{\vphi}\vphi)^2$, and tuning the scalar mass to zero. The action of the deformed theory is given by
	\begin{align}
		S^{\cBcrit}_{k,N}(A_{\mu},\vphi) &= \frac{ik}{4\pi} S_{\rm CS}(A_{\mu}) + S_{b}(A_{\mu},\vphi) + \int \! d^3x \,\frac{\tilde{\lambda}_4}{2N}(\bar{\vphi}\vphi)^2 \ec \label{Scb} \\
		S_{b}(A_{\mu},\vphi) &= \int \! d^3x \, \cD^{\mu}\bar{\vphi}\cD_{\mu}\vphi \ed  \label{Sbb}
	\end{align}
  The theory $\cBreg_{k,N}$ has a regular complex scalar coupled to a gauge field with Chern-Simons interactions, and its action is given by \eqref{Scb} without the double-trace deformation.
  
\item The supersymmetric $\cN=2$ theory with a single chiral multiplet in the fundamental representation will be denoted by $\cT_{k,N}$. In Wess-Zumino gauge, the vector superfield $\cV=\cV^aT^a$ contains the gauge field $A_{\mu}$, gaugino $\lambda$, and two real scalars $\sigma$ and $D$. The chiral superfield $\Phi$ contains a complex scalar $\vphi$, a Dirac fermion $\psi$ and an auxiliary field $F$. The flat-space Euclidean action is given by
	\begin{align}
		S_{k,N}^{\cT}(\Phi,\cV) &= \frac{ik}{4\pi} S_{\rm CS}(\cV) + S_{\mathbf{N}}(\Phi,\cV) \ec \label{Ssusy}\\
		S_{\rm CS}(\cV) &= - \int d^3x\int d^4\theta \int_0^1 \!\!dt\, \tr_N\left[ \cV \bar{D}^{\alpha}\left( e^{2t\cV}D_{\alpha}e^{-2t\cV}\right)\right] \notag\\
		&= \int d^3x \tr_N\left[\epsilon^{\mu\nu\rho}\left(A_{\mu}\partial_{\nu}A_{\rho} - \frac{2i}{3}A_{\mu}A_{\nu}A_{\rho}\right) +2i\sigma D+\bar{\lambda}\lambda\right] \ec \label{Scs}\\
		S_{\mathbf{N}}(\Phi,\cV)  &= \int d^3x \int d^4\theta \, \bar{\Phi} e^{-2\cV} \Phi = \int d^3x \Big[ \cD^{\mu}\bar{\vphi}\cD_{\mu}\vphi - i\bar{\psi}\gamma^{\mu}\cD_{\mu}\psi \notag\\
		&\quad + \bar{\vphi}D\vphi -i\bar{\psi}\sigma\psi +\bar{\vphi}\sigma^2\vphi + i\left(\bar{\vphi}\lambda\psi + \bar{\psi}\bar{\lambda}\vphi\right) - \bar{F}F\Big] \ed \label{SPhi}
	\end{align}
  Here, $D_{\alpha}$ and $\bar{D}_{\alpha}$ are supersymmetric covariant derivatives (see Appendix \ref{conv}).
	After integrating out the auxiliary fields $\sigma$, $D$, $\lambda$ and $F$, the action of $\cT_{k,N}$ becomes
	\begin{align}
		S_{k,N}^{\cT}(A_{\mu},\vphi,\psi) &= \frac{ik}{4\pi}S_{\rm CS}(A_{\mu}) + S_b(A_\mu,\vphi) + S_f(A_\mu,\psi) + S_{bf}(\vphi,\psi) \ec \\
		S_{bf}(\vphi,\psi) &= \int d^3x\left[ -\frac{4\pi i}{k} (\bar{\vphi}\vphi)(\bar{\psi}\psi) + \frac{4\pi^2}{k^2} (\bar{\vphi}\vphi)^3  - \frac{2\pi i}{k} (\bar{\psi}\vphi)(\bar{\vphi}\psi) \right] \ed \label{Sbf}
	\end{align}
\end{itemize}

The actions \eqref{Sf}, \eqref{Scb} and \eqref{Ssusy} formally define non-trivial CFTs in the infrared, and it should be understood that the notation $\cF_{k,N}$, $\cBcrit_{k,N}$ and $\cT_{k,N}$ refers to these CFTs, respectively.
In this paper we will only consider the planar limit obtained by taking $k,N\to\infty$ while keeping the 't Hooft coupling $\lambda=N/k$ fixed. 
In this limit the CFTs are well defined.
For the critical boson theory $\cBcrit_{k,N}$, taking the planar limit means that in practice we compute correlators at finite $\tilde{\lambda}_4$, and then take the limit $\tilde{\lambda}_4\to\infty$ compared to the external momenta, discarding any power-law divergences.\footnote{Here we are implicitly assuming a renormalization scheme such as minimal subtraction with a dimensional reduction regulator, in which tuning the scalar mass to zero along the flow is trivial.} 

Three-dimensional bosonization duality is the conjectured equivalence $\cBcrit_{k,N} \simeq \cF_{\frac{1}{2}-k,|k|-N}$, where $k\in\bZ$, while Giveon-Kutasov duality is the equivalence  $\cT_{k,N}\simeq \cT_{-k,|k|-N+\frac{1}{2}}$ of the $\cN=2$ theory, where now $k\in\bZ+\frac{1}{2}$. 
Planar limit computations are not sensitive to the half-integer shifts in the duality map. We will therefore sometimes omit those shifts in our notations for simplicity.

\subsection{Single-Trace Operators}
\label{ops}

To leading order in the large $N$ expansion, all correlation functions factorize into products of correlators of single-trace operators. We will now review the spectrum of these operators in our theories, focusing on the conformal primaries. In the process, we will provide explicit expressions for all of these operators, and thus fix our normalization conventions for them.

When $\lambda=0$, both the (regular or critical) boson and  fermion theories have one single-trace current for each integer spin $s\geq 1$, which we denote by $J^b_{\alpha_1\cdots\alpha_{2s}}(x)$ and $J^f_{\alpha_1\cdots\alpha_{2s}}(x)$, respectively, where $\alpha_i=1,2$ are spinor indices. It is convenient to suppress the spinor indices of the currents by introducing commuting polarizations $y^{\alpha}$, and defining $J_s(x;y)\equiv y^{\alpha_1}\cdots y^{\alpha_{2s}}J_{\alpha_1\cdots\alpha_{2s}}(x)$. In this notation, the currents in the boson and fermion models can be written explicitly as\footnote{
One can easily verify explicitly that these currents are conserved and traceless, and that they are unique up to an overall normalization. }
\begin{align}
J^b_s &\equiv \sum_{r=0}^s  (-)^{r}\binom{2s}{2r} \partial^r\bar{\vphi} \, \partial^{s-r}\vphi \ec \label{Jb} \\
J^f_s &\equiv y^{\alpha}y^{\beta}\sum_{r=0}^{s-1} (-)^{r+1} \binom{2s}{2r+1} \partial^r\bar{\psi}_{\alpha} \,  \partial^{s-r-1}\psi_{\beta} \ec \label{Jf}
\end{align}
where $\dho\equiv iy^{\alpha}y^{\beta} \gamma^\mu_{\alpha\beta} \partial_{\mu}$. We will always leave the $U(N)$ indices on the fields implicit, it being understood that they are contracted to form $U(N)$ singlets.   When $\lambda\neq  0$ we can make $J^b_s$ and $J^f_s$ in \eqref{Jb} and \eqref{Jf} gauge invariant by simply replacing ordinary derivatives with covariant ones. The currents of spin $s=1$ and $s=2$ correspond to the $U(1)$ current and the stress-tensor, respectively, and are therefore exactly conserved also in the interacting theories. As shown in \cite{Giombi:2011kc,Aharony:2011jz}, the conservation of the currents with $s>2$ is only violated by multi-trace operators, implying that their anomalous dimensions vanish in the planar limit.

In addition, the fermion theory has a scalar operator of dimension $2+O(1/N)$,
\begin{align}
  O_f \equiv - i \bar{\psi} \psi \ec \label{Of}
\end{align}
while the regular scalar theory contains a scalar primary operator of dimension $1+O(1/N)$,
\begin{align}
  \Ob \equiv \bar{\vphi} \vphi \ed \label{Ob}
\end{align}
The critical boson theory has a scalar operator $\widetilde{O}_b$ of dimension $2+O(1/N)$. In order to compute correlators of $\widetilde{O}_b$ by using the Lagrangian description \eqref{Scb}, we insert $\tl_4 \Ob$ and take the IR limit as explained above.

In the $\cN=2$ theory, the set of bosonic single-trace primary operators consists of $O_b$, $O_f$ and all of the currents $J^b_s$ and $J^f_s$.\footnote{In addition, there are fermionic single-trace operators of the form $\bar{\vphi}\partial^s\psi$. We will not consider those in this work.} These operators are packaged into multiplets of the 3d $\cN=2$ superconformal algebra, as we now describe. When $\lambda=0$, the $\cN=2$ theory has a single conserved higher-spin multiplet $\cJ_{\alpha_1\cdots\alpha_{2s}}$ for each integer spin $s\geq 1$.\footnote{See \cite{Nizami:2013tpa} for a recent discussion of conserved higher-spin multiplets in 3d theories with various amount of supersymmetry.} The $\cJ_{\alpha_1\cdots\alpha_{2s}}$ are real superfields that satisfy the conservation constraint $D^{\alpha}\cJ_{\alpha\alpha_2\cdots\alpha_{2s}}=\bar{D}^{\alpha}\cJ_{\alpha\alpha_2\cdots\alpha_{2s}}=0$ on-shell. They can be written in components as
\begin{align}
\cJ_{\alpha_1\cdots\alpha_{2s}} = J_{\alpha_1\cdots\alpha_{2s}} + i\theta^{\beta}\chi_{\beta\alpha_1\cdots\alpha_{2s}} + i\bar{\theta}^{\beta}\bar{\chi}_{\beta\alpha_1\cdots\alpha_{2s}} + \theta^{\beta}\bar{\theta}^{\gamma}\tilde{J}_{\beta\gamma\alpha_1\cdots\alpha_{2s}} + \cdots \ec \label{JsSUSY}
\end{align}
where $J$, $\chi$ and $\tilde{J}$ are conserved currents of spin $s$, $s+\frac{1}{2}$, and $s+1$, respectively. The omitted terms in \eqref{JsSUSY} are determined in terms of these currents by the conservation constraints.

More explicitly, up to an overall constant, the form of the higher-spin superfields \eqref{JsSUSY} (in terms of the chiral superfield) is uniquely fixed and is given by
\begin{align}
\cJ_{s} &\equiv y^{\alpha_1}\cdots y^{\alpha_{2s}}\cJ_{\alpha_1\cdots\alpha_{2s}} \cr
&= \sum_{r=0}^s (-)^r \left[ \binom{2s}{2r} \partial^r \bar{\Phi} \partial^{s-r}\Phi  + \frac{1}{2}\binom{2s}{2r+1} \partial^r \bar{D} \bar{\Phi} \partial^{s-r-1} D \Phi\right] \ec \label{JsSUSYexp}
\end{align} 
where $D\equiv y^{\alpha}D_{\alpha}$ and $\bar{D}\equiv y^{\alpha}\bar{D}_{\alpha}$. By expanding equation \eqref{JsSUSYexp} in the superspace coordinates, the bosonic components of the $\cJ_s$ superfields \eqref{JsSUSY} can be expressed in terms of $J^b_s$ and $J^f_s$. This results in the identifications
\begin{align}
J_s &= J^b_s - J^f_s \ecq \tilde{J}_s = J^b_s + J^f_s \ed \label{JsSUSYbf}
\end{align}

When $\lambda\neq 0$ the $\cJ_s$ are no longer conserved for all $s\geq 1$ (\textit{i.e.}, $D^{\alpha}\cJ_{\alpha\cdots}$ and $\bar{D}^{\alpha}\cJ_{\alpha\cdots}$ are no longer zero). As in the non-supersymmetric theories, one can show that the conservation is only violated by multi-trace operators, implying that the $\cJ_s$ still have canonical dimension in the planar limit \cite{Nizami:2013tpa}. Note that $\cJ_{\alpha\beta}$ is an R-multiplet, whose components include the $U(1)_R$ current $J_{\alpha\beta}$, the supercurrent $\chi_{\alpha\beta\gamma}$ and the  stress-tensor $\tilde{J}_{\alpha\beta\gamma\delta}$, all of which are conserved currents also at finite $N$.\footnote{The canonically normalized R-current is $\frac{1}{2}J_{\alpha\beta}$.} The fact that $D^{\alpha}\cJ_{\alpha\beta}=\bar{D}^{\alpha}\cJ_{\alpha\beta}=0$ is only violated by multi-trace terms implies that $J_{\alpha\beta}$ is, in fact, the exact R-current of the superconformal theory in the planar limit (see also \cite{Safdi:2012re}).

The scalar operators  $O_b$ and $O_f$ are contained in a linear multiplet $\cJ_0$ defined by the condition $D^2\cJ_0=\bar{D}^2\cJ_0=0$. In particular, $\cJ_0=\bar{\Phi}e^{-2\cV}\Phi$, and after integrating out the auxiliary fields it can be written in components as
\begin{align}
\cJ_0 &= O_b + i \theta\bar{\theta} \left(O_f + \frac{4\pi}{k}O_b^2\right) + \bar{\theta}\gamma^{\mu}\theta \tilde{J}_{\mu} + \cdots \ec \label{cJ0} \\ 
\tilde{J}_{\mu} &= J^b_{\mu}+J^f_{\mu} = i
(\bar{\varphi} \cD_\mu \varphi - \cD_\mu \bar{\varphi} \cdot \varphi)
- \bar{\psi}\gamma_{\mu}\psi \ed \label{tJf}
\end{align}
Note that $\tilde{J}_{\mu}$ is a conserved flavor current, and therefore $\cJ_0$ has dimension $\Delta=1$ for all $N$ and $k$. 

\subsection{Relations Between 2-Point Correlators}
\label{cptPrel}

Some planar correlators of single-trace operators in the $\cN=2$ theory can be written in terms of correlators in the non-supersymmetric theories.
In this section we explain these relations, which are used extensively in this work.

\begin{figure}
  \centering
  \includegraphics[width=0.6\textwidth]{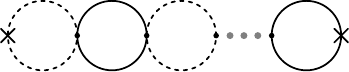}
  \caption{A generic planar contribution to $\langle \Ob \Of \rangle_\cT$.
  Dashed lines represent scalar propagators, solid lines represent fermion propagators, and crosses denote the operator insertions.
  }
  \label{fig:ObOf}
\end{figure}

As a basic example, consider the planar 2-point function $\langle \Ob \Of \rangle_{\cT}$ in the supersymmetric theory. This correlator vanishes at separated points, because $\Ob$ and $\Of$ have different conformal dimensions. However, $\langle \Ob \Of \rangle_{\cT}$ is not zero in momentum-space, because it contains a non-trivial contact term. The planar contributions to this correlator take the form shown in Figure \ref{fig:ObOf}.
Each matter loop can include any planar configuration of gauge bosons, which are not shown explicitly.
Therefore, the momentum-space correlator in the supersymmetric theory can be written in terms of correlators of the regular bosonic and fermionic theories as follows.
\begin{align}
  \langle \Ob \Of \rangle_{\cT_{k,N}} &=
  - \frac{k}{4\pi}
  \sum_{n=1}^\infty \left[ \frac{(4\pi)^2}{k^2}
  \langle O_b O_b \rangle_{\cBreg_{k,N}} \langle O_f O_f \rangle_{\cF_{k,N}}
  \right]^n 
  \ed \label{ObOf}
\end{align}
The correlator can be computed explicitly using the known results for planar 2-point functions \cite{Aharony:2012nh,GurAri:2012is}, but we will not require its explicit form.
We see that contributions to $\langle O_b O_f \rangle_{\cT}$ factorize through the double-trace $(\bar{\varphi} \varphi) (\bar{\psi} \psi)$ vertex.
Other planar correlators of $\Ob$, $\Of$ and of the currents in the supersymmetric theory have similar contributions, that factorize through the multi-trace vertices of the supersymmetric theory.
We claim that the relation \eqref{ObOf} is not affected by renormalization, so it is an exact relation of the continuum theories in the planar limit.
This follows from the fact that the theories involved are not renormalized in the planar limit: there are no logarithmic divergences, and therefore no need to introduce counter-terms for either the couplings or for the operators \cite{Aharony:2011jz}.

\begin{figure}
  \centering
  \includegraphics[width=0.6\textwidth]{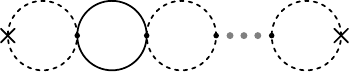}
  \caption{A generic planar contribution to $\langle \Ob \Ob \rangle_\cT$.
  The notation is the same as in Figure \ref{fig:ObOf}.
  }
  \label{fig:ObOb}
\end{figure}

Next, consider the correlator $\langle \Ob \Ob \rangle_{\cT}$.
A typical planar contribution is shown in Figure \ref{fig:ObOb}, and we can write this correlator as
\begin{align}
  \langle \Ob \Ob \rangle_{\cT_{k,N}} &=
  \langle \Ob \Ob \rangle_{\cBreg_{k,N}} \left[ 1 - 
  \frac{k}{4\pi} \langle \Ob \Of \rangle_{\cT_{k,N}}
  \right] \ed
  \label{ObOb}
\end{align}
The value of the contact term $\langle \Ob \Of \rangle_{\cT}$ can be shifted by introducing a conformally-invariant term in the action of the form $\int \! d^3x \, \hat{D} \hat{\sigma}$, where $\hat{D}$ and $\hat{\sigma}$ source $\Ob$ and $\Of$ respectively.
Introducing this term would invalidate the equality \eqref{ObOb} because this term does not affect $\langle \Ob \Ob \rangle_{\cT}$.
Therefore, for the rest of this section we set this term to zero, as well as any other finite counter-term that can affect correlators at coincident points. 

The same idea can be used to compute correlators in the critical bosonic theory, where contributions factorize through the double-trace vertex $(\bar{\varphi} \varphi)^2$.
The critical bosonic theory has a scalar operator $\tOb$ of dimension $2 + O(1/N)$.
As explained above, we compute correlators of this operator by flowing to it from $\tilde{\lambda}_4 \Ob$.
For example, to compute the 2-point function we consider the correlator
\begin{align}
  \tl_4^2 \, \la \Ob(p) \Ob \ra_{\cBcrit_{k,N}} &=
  \tl_4^2 \, \la \Ob(p) \Ob \ra_{\cBreg_{k,N}}
  \sum_{n=0}^\infty \left( - \frac{\tl_4}{N} \la \Ob(p) \Ob \ra_{\cBreg_{k,N}} \right)^n
  \ed
\end{align}
Taking the IR limit $\tl_4^{-1} |p| \to 0$ and discarding a linear divergence, we find that
\begin{align}
  \la \tOb \tOb(p) \ra_{\cBcrit_{k,N}} &= 
  - \frac{N^2}{\la \Ob \Ob(p) \ra_{\cBreg_{k,N}}}
  \ed \label{OtOt}
\end{align}
Using the known result of the bosonic 2-point function, we find that
\begin{align}
  \la \tOb \tOb(p) \ra_{\cBcrit_{k,N}} &= 
  - \frac{4\pi \lambda N}{\tan ( \pi\lambda / 2 )} |p| \ed
\end{align}

Next, let us consider 2-point functions of currents.
For a fermionic current $\Jf_s$ in the supersymmetric theory, perturbation theory in the multi-trace couplings tells us that
\begin{align}
  \langle \Jf_{s}(p) \Jf_{s} \rangle_{\cT_{k,N}}
  &=
  \langle \Jf_{s}(p) \Jf_{s} \rangle_{\cF_{k,N}}
  + 
  \left( \frac{4\pi}{k} \right)^2 
  \langle \Jf_{s}(p) O_f \rangle_{\cF_{k,N}} \cdot
  \langle O_b(p) O_b \rangle_{\cT_{k,N}} \cdot
  \langle O_f(p) \Jf_{s} \rangle_{\cF_{k,N}} \ed
  \label{JJ}
\end{align}
Here, the currents have arbitrary polarizations, which we do not write explicitly to avoid clutter.
We would now like to argue that the correlator $\langle \Jf_s \Of \rangle$ that appears on the right-hand of \eqref{JJ} vanishes.
This may seem obvious due to conformal symmetry, but this argument only applies to the correlator at separated points.
If $\langle \Jf_s \Of \rangle$ contains a non-vanishing contact term, which is equivalent to a polynomial in the momentum, then this term would contribute to $\langle \Jf_{s} \Jf_{s} \rangle_{\cT}$ \textit{even at separated points}.
This is because the overall term on the right-hand side of \eqref{JJ} would not be a contact term in this case.
In Appendix \ref{2ptapp} we prove that all planar correlators of the form $\langle J O \rangle$, with one current and one scalar operator insertion, vanish in our theories even at coincident points.
We therefore have the relation
\begin{align}
  \langle \Jf_s \Jf_s \rangle_{\cT_{k,N}} &=
  \langle \Jf_s \Jf_s \rangle_{\cF_{k,N}} \ed
\end{align}
A similar argument for bosonic currents leads to the following equalities for planar 2-point functions in Chern-Simons vector models.
\begin{align}
  \langle \Jb_s \Jb_s \rangle_{\cT_{k,N}} &=
  \langle \Jb_s \Jb_s \rangle_{\cBcrit_{k,N}} =
  \langle \Jb_s \Jb_s \rangle_{\cBreg_{k,N}} \ec \cr
  \langle \Jf_s \Jb_s \rangle_{\cT_{k,N}} &= 0
  \ed \label{basic2pt}
\end{align}
These relations are exact in the planar limit, and hold even at coincident points (\textit{i.e.} including contact terms).
Similar relations for 3-point functions will be derived in Section \ref{cpt}.

\section{Duality Map of the $\cN=2$ Theory}
\label{N2duality}

Our goal in this section is to determine how the single-trace conformal primary operators of the $\cN=2$ theory $\cT_{k,N}$ map under Giveon-Kutasov duality. As we saw in Section \ref{setup}, for each spin $s$ the theory $\cT_{k,N}$ has two single-trace conformal primaries $\Jb_s$ and $\Jf_s$, but only one single-trace superconformal multiplet $\cJ_s$. What we will determine is how $\Jb_s$ and $\Jf_s$ mix under the duality. 

Let us first briefly summarize how the global symmetry charges of $\cT_{k,N}$  transform under Giveon-Kutasov duality \cite{Giveon:2008zn,Benini:2011mf}. The theory has two global $U(1)$ symmetries, one of which is the flavor symmetry generated by $\tilde{J}_{\mu}=J^b_{\mu}+J^f_{\mu}$, under which $\Phi$ has charge $1$. The other $U(1)$ is an R-symmetry generated by $\frac{1}{2}J_{\mu} = \frac{1}{2}\left(J^b_{\mu}-J^f_{\mu}\right)$, under which $\vphi$ has charge $r=\frac{1}{2}$ and the gaugino $\lambda$ has charge $1$.\footnote{The topological $U(1)$ symmetry generated by $J_{\mu}^{\rm top.}=\frac{ik}{8\pi}\varepsilon_{\mu\nu\rho}\tr(F^{\mu\nu})$ is equivalent to the flavor current $\tilde{J}_{\mu}$ by the equations of motion.} As we discussed in Section \ref{ops}, in the planar limit this is the exact R-current of the SCFT $\cT_{k,N}$. 

Let $\cT_{k_e,N_e}$ denote the `electric' $\cN=2$ theory. Its `magnetic' dual is given by a $U(N_m)_{k_m}$ Chern-Simons gauge theory coupled to a chiral superfield $\widetilde{\Phi}$ in the anti-fundamental representation, with the identifications 
\begin{align}
k_m=-k_e \ecq N_m=|k_e|+\frac{1}{2}-N_e \ed
\end{align} 
Moreover, $\widetilde{\Phi}$ has charge $-1$ under the flavor $U(1)$ and the R-charges are the same as in the electric theory (in particular, $\widetilde{\Phi}$ has R-charge $1-r = \frac{1}{2}$). By a suitable field redefinition, the magnetic theory can be written in terms of a chiral multiplet in the fundamental representation (with the same global symmetry charges as above); namely, it can be written as the theory $\cT_{k_m,N_m}$.

\subsection{Map of Single-Trace Operators}
\label{mst}

Let us now deduce the duality map of single-trace operators in the theory $\cT_{k,N}$. As we saw in Section \ref{ops}, the theory $\cT_{k,N}$ has one multiplet $\cJ_s$ for each integer spin ($s=0,1,2,\ldots$). $\cJ_s$ includes single-trace operators, plus possible multi-trace corrections that are not important for us. The dual theory $\cT_{-k,|k|-N+1/2}$ has the same spectrum of single-trace multiplets. Because there is only one single-trace multiplet of each spin, under the duality $\cT_{k,N}\to\cT_{-k,|k|-N+1/2}$ the $\cJ_s$ multiplets must map to themselves. We will now determine the overall numerical factors that can appear in the transformation
\begin{align}
\cJ_s \to c_s \cJ_s \ed \label{cJmap}
\end{align}
Note that $|c_s|$ depends on the overall normalization of the $\cJ_s$ that was defined in Section \ref{setup}, but there can also be signs in the duality map. In fact we will show that, in our normalization, $\cJ_s \to (-)^{s+1} \cJ_s$ under Giveon-Kutasov duality, and in components
\begin{align}
\Ob \to -\Ob \ecq \Of \to -\Of \ecq \Jb_s \to (-)^s \Jf_s 
\ecq \Jf_s \to (-)^s \Jb_s 
\ed \label{TknOpMap}
\end{align}
The derivation is slightly technical and can be safely skipped by the reader.
 
The multiplets $\cJ_0$ and $\cJ_1$ transform according to the duality maps of the flavor and R symmetries, respectively, which implies that $c_0=-1$ and $c_1=1$.\footnote{ The mapping of the flavor $U(1)$ multiplet $\cJ_0$ is exact, but that of the R-multiplet $\cJ_1$ is only valid in the planar limit. At next-to-leading order in $1/N$ the flavor and R-current can mix, $\langle J_{\mu} \tilde{J}_{\mu} \rangle = O(1/N)$, and the multiplet we call $\cJ_1$ is then not the superconformal R-multiplet.} In particular, the duality map of the components $O_b$, $O_f$, $J^b_1$ and $J^f_1$, of the superfields $\cJ_0$ and $\cJ_1$ (see \eqref{JsSUSY}, \eqref{JsSUSYexp} and \eqref{cJ0}) is determined to be
\begin{align}
\Ob   \to -\Ob   \ecq     \Of \to -\Of\ecq \Jb_1 \to -\Jf_1 \ecq   \Jf_1 &\to -\Jb_1 \ed \label{basis}
\end{align}

In determining the remaining constants $c_s$ it is useful to consider the duality map in the basis of $J^b_s$ and $J^f_s$. Let $\cM_s$ be the 2-by-2 duality transformation matrix defined by
\begin{align}
\begin{pmatrix}
\Jb_s \\ \Jf_s
\end{pmatrix} 
\rightarrow 
\cM_s \cdot \begin{pmatrix}
\Jb_s \\ \Jf_s
\end{pmatrix} \ed
\end{align}
To determine $\cM_s$, first note that because $\cJ_s$ with different $s$ values do not mix we have\footnote{The correlators we are considering in this section are all at separated points.}
\begin{align}
  \langle \cJ_{s-1} \, \cJ_{s} \rangle = O(1/N) \quad \Rightarrow \quad 
  \left. \langle \cJ_{s-1} \, \cJ_{s} \rangle \right|_{\theta\bar{\theta}} =
  \langle J_s \, \tilde{J}_s\rangle = O(1/N) \ed \label{JJt}
\end{align}
Plugging into \eqref{JJt} the expression \eqref{JsSUSYexp} of $J_s$ and $\tilde{J}_s$ in terms of $J^b_s$ and $J^f_s$, we obtain
\begin{align}
  \langle J^b_s J^b_s \rangle_{\cT_{k,N}} = 
  \langle J^f_s J^f_s \rangle_{\cT_{k,N}} + O(1/N) \ed \label{2pntMap}
\end{align}
Using also the fact that $\langle J^{b\vphantom{f}}_s J^f_s \rangle_{\cT_{k,N}} = O(1/N)$, which is easy to prove diagrammatically, we conclude that the matrix of planar 2-point functions of $\Jb_s$ and $\Jf_s$ is proportional to the identity. Therefore, the transformation matrix $\cM_s$ must be proportional to an orthogonal matrix in order to preserve the matrix of 2-point functions.

The rest of the argument follows by induction, whose basis is given in \eqref{basis}. Assume we have already determined that $\Jb_{s-1}\to (-)^{s-1} \Jf_{s-1}$ and $\Jf_{s-1}\to (-)^{s-1} \Jb_{s-1}$, for some $s \geq 2$. In particular, $\cJ_{s-1}\to (-)^s\cJ_{s-1}$,  implying that $\tilde{J}_s \to (-)^{s} \tilde{J}_s$. Going to the $\Jb_s$, $\Jf_s$ basis we conclude that $\cM_s^T$ has an eigenvector $(1,1)$ with eigenvalue $(-)^s$. To summarize, we learned that
\begin{align}
	\cM_s \cdot \cM_s^T \propto 1 \ecq \cM_s^T \cdot \begin{pmatrix}
	1\\ 1
	\end{pmatrix} = (-)^{s}\begin{pmatrix}
	1\\ 1
	\end{pmatrix} \ed \label{Deq}
\end{align}

The equations \eqref{Deq} have two solutions for the duality map, given by
\begin{align}
\mathbf{1}) \,\,\, \Jb_s &\to (-)^s \Jf_s \ecq \Jf_s\to (-)^s \Jb_s \ec \label{sol1} \\ 
\mathbf{2}) \,\,\, \Jb_s &\to(-)^s\Jb_s\ecq \Jf_s\to(-)^s\Jf_s \ed \label{sol2}
\end{align}
The second solution is readily seen to be inconsistent with the duality. Indeed, using \eqref{basis} it implies that in the planar limit we have the identities
\begin{align}
  \langle \Jb_{1}\Jb_{1}\Jb_{s} \rangle_{\cT_{k,N}} &= (-)^s \langle \Jf_{1}\Jf_{1}\Jb_{s} \rangle_{\cT_{-k,|k|-N}} \ec \label{inc1}\\
  \langle \Ob \Jb_{1} \Jb_{s} \rangle_{\cT_{k,N}} &= (-)^s \langle \Ob \Jf_{1}\Jb_{s} \rangle_{\cT_{-k,|k|-N}} \ed \label{inc2}
\end{align}
In a free scalar theory, $\langle J_{s_1}^b J_{s_2}^b J_{s_3}^b \rangle$ is non-zero if and only if $s_1+s_2+s_3=$ even (see e.g., \cite{Giombi:2011rz}).
Let us turn on a weak coupling, so that both the theory and its dual are interacting. If $s$ is even then the left-hand side of \eqref{inc1} is still non-zero in the weakly-coupled theory, while the right-hand side is identically zero in the planar limit (c.f. equation \eqref{basic3pt} below). If $s$ is odd then we reach the same conclusion by considering equation \eqref{inc2}. This concludes the derivation of the mapping \eqref{TknOpMap}.

\section{Bosonization from Perturbation Theory}
\label{cpt}

In this section we show that, for a large class of planar correlators, the supersymmetric duality and non-supersymmetric bosonization dualities are equivalent.
For this purpose we use perturbation theory in the multi-trace couplings of the $\cN=2$ theory, as explained in Section \ref{cptPrel}.
We will first prove this statement for all 3-point functions, and then extend the proof to the 4-point function of spin 1 currents, all at separated points.

\subsection{3-Point Functions}

The supersymmetric duality implies the following relations.
\begin{align}
  \langle \Jb_{s_1} \Jb_{s_2} \Jb_{s_3} \rangle_{\cT_{k_e,N_e}} &=
  (-)^{s_1+s_2+s_3}
  \langle \Jf_{s_1} \Jf_{s_2} \Jf_{s_3} \rangle_{\cT_{k_m,N_m}} 
  \ec  \\
  \langle \Jb_{s_1} \Jb_{s_2} \Ob \rangle_{\cT_{k_e,N_e}} &=
  (-)^{s_1 + s_2 + 1} 
  \langle \Jf_{s_1} \Jf_{s_2} \Ob \rangle_{\cT_{k_m,N_m}} 
  \label{JJOT} \ec \\
  \langle \Jb_{s_1} \Ob \Ob \rangle_{\cT_{k_e,N_e}} &=
  (-)^{s_1} 
  \langle \Jf_{s_1} \Ob \Ob \rangle_{\cT_{k_m,N_m}} \ed
\end{align}
We will now prove that they are equivalent to the non-supersymmetric bosonization relations
\begin{align}
  \langle \Jb_{s_1} \Jb_{s_2} \Jb_{s_3} \rangle_{\cBcrit_{k_e,N_e}} &=
  (-)^{s_1+s_2+s_3}
  \langle \Jf_{s_1} \Jf_{s_2} \Jf_{s_3} \rangle_{\cF_{k_m,N_m}} 
  \ec \label{JJJ} \\
  \langle \Jb_{s_1} \Jb_{s_2} \tOb \rangle_{\cBcrit_{k_e,N_e}} &=
  (-)^{s_1 + s_2} \left( -4\pi\lambda_e \right)
  \langle \Jf_{s_1} \Jf_{s_2} \Of \rangle_{\cF_{k_m,N_m}} 
  \ec \label{JJO} \\
  \langle \Jb_{s_1} \tOb \tOb \rangle_{\cBcrit_{k_e,N_e}} &=
  (-)^{s_1} \left( 4\pi \lambda_e \right)^2
  \langle \Jf_{s_1} \Of \Of \rangle_{\cF_{k_m,N_m}} \label{JOO} \ed
\end{align}
These hold in the planar limit at separated points, for any positive spins $s_1,s_2,s_3$.\footnote{
The correlator $\langle \tOb\tOb  \tOb \rangle_{\cBcrit}$ and its fermionic counterpart are pure contact terms, and will not be considered here.}
At the level of 3-point functions, the relations above imply the following mapping of operators between the bosonic theory $\cB^{\mathrm{crit.}}_{k_e,N_e}$ and the fermionic theory $\cF_{k_m,N_m}$:
\begin{align}
  \tOb \to -4\pi \lambda_e \Of \ecq \Jb_s \to (-)^s \Jf_s  \ed \label{Otmap}
\end{align}
The minus signs in the duality map of the currents were not noticed previously; they are consistent with all the explicit computations of correlation functions that were done in the past, as those particular correlators were not sensitive to those signs.

We begin with the 3-point function of fermionic currents, which can be written as
\begin{align}
  \langle \Jf_{s_1} \Jf_{s_2} \Jf_{s_3} \rangle_{\cT_{k,N}} &=
  \langle \Jf_{s_1} \Jf_{s_2} \Jf_{s_3} \rangle_{\cF_{k,N}} +
  \left( \frac{4\pi}{k} \right)^2
  \langle \Jf_{s_1} \Of \rangle_{\cF_{k,N}}
  \langle \Ob \Ob \rangle_{\cT_{k,N}}
  \langle \Of \Jf_{s_2} \Jf_{s_3} \rangle_{\cF_{k,N}} +
  \cdots \ec
  \label{J3}
\end{align}
where the remaining terms all include factors of $\langle \Jf_s \Of \rangle_{\cF_{k,N}}$.
We show in Appendix \ref{2ptapp} that these 2-point functions vanish (also at coincident points), so we find an equality between the 3-point functions in the supersymmetric and fermionic theories.
(As explained in Section \ref{cptPrel}, factorization relations such as \eqref{J3} hold when all the finite counter-terms that affect correlators at coincident points are set to zero.)
Extending this argument to other 3-point functions, we find the relations
\begin{align}
  \langle \Jf_{s_1} \Jf_{s_2} \Jf_{s_3} \rangle_{\cT_{k,N}} &=
  \langle \Jf_{s_1} \Jf_{s_2} \Jf_{s_3} \rangle_{\cF_{k,N}} \ec \cr
  \langle \Jb_{s_1} \Jb_{s_2} \Jb_{s_3} \rangle_{\cT_{k,N}} &=
  \langle \Jb_{s_1} \Jb_{s_2} \Jb_{s_3} \rangle_{\cBcrit_{k,N}} =
  \langle \Jb_{s_1} \Jb_{s_2} \Jb_{s_3} \rangle_{\cBreg_{k,N}} \ec \cr
  \langle \Jf_{s_1} \Jb_{s_2} \Jb_{s_3} \rangle_{\cT_{k,N}} &= 0 \ec \cr
  \langle \Jf_{s_1} \Jf_{s_2} \Jb_{s_3} \rangle_{\cT_{k,N}} &= 0
  \ed \label{basic3pt}
\end{align}
These relations hold for any positive spins $s_1,s_2,s_3$.
The duality of the supersymmetric theory implies that $\langle \Jf_{s_1} \Jf_{s_2} \Jf_{s_3} \rangle_{\cT_{k_e,N_e}} = (-)^{s_1+s_2+s_3} \langle \Jb_{s_1} \Jb_{s_2} \Jb_{s_3} \rangle_{\cT_{k_m,N_m}}$, and the equality \eqref{JJJ} follows.

Next, consider correlators with two bosonic currents of spins $s_1,s_2$ and with one scalar insertion.
In the supersymmetric theory, we can write these as
\begin{align}
  \langle \Jb_{s_1} \Jb_{s_2} \Ob(p) \rangle_{\cT_{k_e,N_e}} &=
  \left[ 1 - \frac{4\pi}{k_e} \langle \Ob \Of \rangle_{\cT_{k_e,N_e}} \right] 
  \langle \Jb_{s_1} \Jb_{s_2} \Ob(p) \rangle_{\cBreg_{k_e,N_e}} 
  \cr &=
  \left[ 1 - \frac{4\pi}{k_e} \langle \Ob \Of \rangle_{\cT_{k_e,N_e}} \right] 
  \frac{1}{N_e}
  \langle \Ob \Ob(p) \rangle_{\cBreg_{k_e,N_e}} 
  \langle \Jb_{s_1} \Jb_{s_2} \tOb(p) \rangle_{\cBcrit_{k_e,N_e}} 
  \cr &=
  \frac{1}{N_e}
  \langle \Ob \Ob(p) \rangle_{\cT_{k_e,N_e}} 
  \langle \Jb_{s_1} \Jb_{s_2} \tOb(p) \rangle_{\cBcrit_{k_e,N_e}} 
  \ed
  \label{elecSide}
\end{align}
In the second line we used the following relation between the regular and critical bosonic theories.
\begin{align}
  \langle \Jb_{s_1} \Jb_{s_2} \tOb(p) \rangle_{\cBcrit_{k,N}} &=
  \lim_{\tilde{\lambda}_4 \to \infty}
  \tilde{\lambda}_4
  \langle \Jb_{s_1} \Jb_{s_2} \Ob(p) \rangle_{\cBreg_{k,N}} 
  \cdot \sum_{n=0}^\infty
  \left[ - \frac{\tilde{\lambda}_4}{N}
  \langle \Ob \Ob(p) \rangle_{\cBreg_{k,N}}
  \right]^n \cr
  &= \frac{N
  \langle \Jb_{s_1} \Jb_{s_2} \Ob(p) \rangle_{\cBreg_{k,N}} }
  { \langle \Ob \Ob(p) \rangle_{\cBreg_{k,N}} } \ed
  \label{JJOrel}
\end{align}
The supersymmetric duality \eqref{JJOT} implies that the correlator in \eqref{elecSide} is equal to\footnote{
The relations in this section are all correct up to $O(1/N)$ corrections; $O(1)$ shifts of the level are ignored.
}
\begin{align}
  (-)^{s_1 + s_2 + 1}
  \langle \Jf_{s_1} \Jf_{s_2} \Ob(p) \rangle_{\cT_{k_m,N_m}} &=
  (-)^{s_1 + s_2+1} \frac{4\pi}{k_e}
  \langle \Jf_{s_1} \Jf_{s_2} \Of(p) \rangle_{\cF_{k_m,N_m}} 
  \langle \Ob \Ob(p) \rangle_{\cT_{k_m,N_m}} 
  \ed \label{magSide}
\end{align}
This proves the bosonization relation \eqref{JJO} (notice that $\langle \Ob \Ob \rangle_{\cT}$ is invariant under the duality). A similar calculation proves \eqref{JOO}.
Note that we did not need to use the explicit form of the 2-point functions.

\subsection{4-Point Function of Spin 1 Currents}

\begin{figure}
  \centering
  \includegraphics[width=0.8\textwidth]{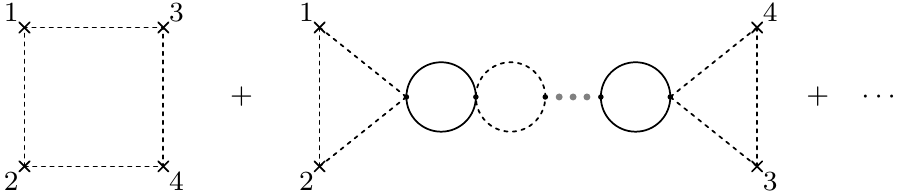}
  \caption{Planar diagrams contributing to the 4-point function of bosonic spin-1 currents in the $\cN=2$ theory.
  The notation is the same as in Figure \ref{fig:ObOf}.
  }
  \label{fig:JJJJ}
\end{figure}

In this section we prove that the supersymmetric duality relation for the 4-point function of spin 1 currents,
\begin{align}
  \langle \Jb_1 \Jb_1 \Jb_1 \Jb_1 \rangle_{\cT_{k_e,N_e}} =
  \langle \Jf_1 \Jf_1 \Jf_1 \Jf_1 \rangle_{\cT_{k_m,N_m}} \ec
  \label{JJJJsusy}
\end{align}
implies the non-supersymmetric duality
\begin{align}
  \langle \Jb_1 \Jb_1 \Jb_1 \Jb_1 \rangle_{\cBcrit_{k_e,N_e}} =
  \langle \Jf_1 \Jf_1 \Jf_1 \Jf_1 \rangle_{\cF_{k_m,N_m}} \ed
  \label{JJJJ}
\end{align}
The supersymmetric 4-point function can be written as
\begin{align}
  \langle \Jb_1(p_1) \Jb_1(p_2) &\Jb_1(p_3) \Jb_1(p_4) \rangle_{\cT_{k_e,N_e}} =
  \cr &\quad
  \langle \Jb_1(p_1) \Jb_1(p_2) \Jb_1(p_3) \Jb_1(p_4) \rangle_{\cBreg_{k_e,N_e}} 
  + \cr &\quad
  \left( \frac{4\pi}{k_e} \right)^2
  \langle \Jb_1(p_1) \Jb_1(p_2) \Ob \rangle_{\cBreg_{k_e,N_e}} 
  \langle \Jb_1(p_3) \Jb_1(p_4) \Ob \rangle_{\cBreg_{k_e,N_e}} 
  \langle \Of \Of(p_1+p_2) \rangle_{\cT_{k_e,N_e}}
  \cr &\quad
  + (\mathrm{two~permutations}) \ed
  \label{Jb4}
\end{align}
This is shown diagramatically in Figure \ref{fig:JJJJ}.

On the magnetic side, we have
\begin{align}
  \langle \Jf_1(p_1) \Jf_1(p_2) &\Jf_1(p_3) \Jf_1(p_4) \rangle_{\cT_{k_m,N_m}} =
  \cr &\quad
  \langle \Jf_1(p_1) \Jf_1(p_2) \Jf_1(p_3) \Jf_1(p_4) \rangle_{\cF_{k_m,N_m}} +
  \cr &\quad
  \left( \frac{4\pi}{k_m} \right)^2
  \langle \Jf_1(p_1) \Jf_1(p_2) \Of \rangle_{\cF_{k_m,N_m}} 
  \langle \Jf_1(p_3) \Jf_1(p_4) \Of \rangle_{\cF_{k_m,N_m}} 
  \langle \Ob \Ob(p_1+p_2) \rangle_{\cT_{k_m,N_m}}
  \cr &\quad
  + (\mathrm{two~permutations}) \ed
  \label{Jf4}
\end{align}
Using \eqref{JJO} and \eqref{JJOrel}, we can write the 3-point function as
\begin{align}
  \langle \Jf_1 \Jf_1 \Of(p) \rangle_{\cF_{k_m,N_m}} &=
  - \frac{1}{4\pi\lambda_e}
  \langle \Jb_1 \Jb_1 \tOb(p) \rangle_{\cBcrit_{k_e,N_e}}
  =
  - \frac{k_e}{4\pi}
  \frac{\langle \Jb_1 \Jb_1 \Ob(p) \rangle_{\cBreg_{k_e,N_e}}}{
  \langle \Ob \Ob(p) \rangle_{\cBreg_{k_e,N_e}}}
  \ed
\end{align}
This holds true at separated points, but in our derivation it will be important that it is also true at coincident points.
The reason for this was explained in the discussion below equation \eqref{JJ}: a scheme dependent contact term in the 3-point function (\textit{i.e.} a polynomial in the momenta) will affect the 4-point functions \eqref{Jb4} and \eqref{Jf4} \emph{even at separated points}.
It is therefore important that all such terms map correctly under bosonization.
We will prove that this is indeed the case below.

Continuing with this assumption, equation \eqref{JJJJsusy} implies that
\begin{align}
  \langle \Jf_1(p_1) \Jf_1(p_2) &\Jf_1(p_3) \Jf_1(p_4) \rangle_{\cF_{k_m,N_m}} 
  =
  \cr &\quad
  \langle \Jb_1(p_1) \Jb_1(p_2) \Jb_1(p_3) \Jb_1(p_4) \rangle_{\cBreg_{k_e,N_e}} 
  + \cr &\quad
  \langle \Jb_1(p_1) \Jb_1(p_2) \Ob \rangle_{\cBreg_{k_e,N_e}} 
  \langle \Jb_1(p_3) \Jb_1(p_4) \Ob \rangle_{\cBreg_{k_e,N_e}} 
  \times \cr &\quad \qquad \qquad
  \Bigg[
  \left( \frac{4\pi}{k_e} \right)^2
  \langle \Of \Of(p_1+p_2) \rangle_{\cT_{k_e,N_e}}
  - 
  \frac{\langle \Ob \Ob(p_1+p_2) \rangle_{\cT_{k_m,N_m}}}{
  \langle \Ob \Ob(p_1+p_2) \rangle^2_{\cBreg_{k_e,N_e}}}
  \Bigg]
  \cr &\quad
  + (\mathrm{two~permutations}) \ed
  \label{JJJJ1}
\end{align}
Now, consider the critical bosonic theory.
The 4-point function can be written as
\begin{align}
  \langle \Jb_1(p_1) \Jb_1(p_2) &\Jb_1(p_3) \Jb_1(p_4) \rangle_{\cBcrit_{k_e,N_e}} =
  \cr &\quad
  \langle \Jb_1(p_1) \Jb_1(p_2) \Jb_1(p_3) \Jb_1(p_4) \rangle_{\cBreg_{k_e,N_e}} 
  + \cr &\quad
  N_e^{-2}
  \langle \Jb_1(p_1) \Jb_1(p_2) \Ob \rangle_{\cBreg_{k_e,N_e}} 
  \langle \Jb_1(p_3) \Jb_1(p_4) \Ob \rangle_{\cBreg_{k_e,N_e}} 
  \langle \tOb \tOb(p_1+p_2) \rangle_{\cBcrit_{k_e,N_e}}
  \cr &\quad
  + (\mathrm{two~permutations}) \ed
  \label{JJJJ2}
\end{align}
It is easy to check that the right-hand sides of \eqref{JJJJ1} and \eqref{JJJJ2} are equal, and this proves the bosonization relation \eqref{JJJJ}.\footnote{
This follows from the relations
\begin{align}
  \la \Of \Of \ra_{\cT_{k_e,N_e}} &=
  - \frac{k_e}{4\pi} \frac{\la \Of \Ob \ra_{\cT_{k_e,N_e}}}
  {\la \Ob \Ob \ra_{\cBreg_{k_e,N_e}}} \ec \cr
  \la \Ob \Ob \ra_{\cT_{k_m,N_m}} &=
  \la \Ob \Ob \ra_{\cT_{k_e,N_e}} =
  \la \Ob \Ob \ra_{\cBreg_{k_e,N_e}} 
  \left[ 1 - \frac{4\pi}{k_e} \la \Of \Ob \ra_{\cT_{k_e,N_e}} \right]
  \ec
\end{align}
and from equation \eqref{OtOt}.
}
Note that, again, we did not need to use any explicit expressions for the 2-point functions, but only simple relations that follow from perturbation theory in the multi-trace couplings.

It is left to show that the 3-point function $\langle \Jf_1 \Jf_1 \Of \rangle$ maps correctly under the bosonization, including contact terms.
We assume that universal (scheme-independent) contact terms agree under the duality, because such terms are physical observables.
On the other hand, scheme-dependent contact terms (which correspond to polynomials in the momenta) can be shifted by local counter-terms that are composed of the background fields.
Such contact terms might not agree under the duality unless we tune the corresponding counter-terms, but we had already set all such counter-terms to zero.
In other words, if we find scheme-dependent contact terms whose value does not map correctly under the duality, then our argument does not go through.
The only scheme-dependent contact term that we can write down in $\langle \Jf_\mu \Jf_\nu \Of \rangle$ is $\delta_{\mu\nu}$. 
This term is ruled out because it is not conserved.

\subsection{Other Correlators}

It is plausible that the argument above can be generalized to other higher-point functions.
The argument for 4-point functions of currents with general spins goes through as-is, except that one must now prove that the 3-point functions of the form $\langle J J O \rangle$ have no scheme-dependent contact terms (or that such contact terms, if they exist, map correctly under the duality).
For other correlators such as $\langle JJJO \rangle$, or for 5-point functions and above, perturbation theory in the multi-trace couplings becomes more cumbersome.
One complication is that, for most correlators, both the $(\bar{\varphi} \varphi) (\bar{\psi} \psi)$ and the $(\bar{\varphi} \varphi)^3$ vertices of the supersymmetric theory appear in the factorization.
Instead of following this route, in the next section we will present an argument that proves the bosonization of all planar correlators.

\section{Bosonization from Double-Trace Flow}
\label{dualityProof}

In this section we will give a simpler derivation of the basic bosonization duality from the Giveon-Kutasov duality, which applies to all planar correlators in those theories. To make contact with the non-supersymmetric theories we add a  $(\bar{\vphi}\vphi)^2$ double-trace deformation to the $\cN=2$ action and flow to a CFT in the IR. The induced duality of that non-supersymmetric CFT will be shown to imply the duality $\cB^{\mathrm{crit.}}_{k, N} \simeq \cF_{-k,|k|-N}$. The   double-trace deformation $(\bar{\vphi}\vphi)^2$ is equivalent to coupling the flavor $U(1)$ current to a background vector superfield, and making the top component of the latter dynamical. We therefore start in Section \ref{bgvector} by reviewing how to carefully couple both sides of Giveon-Kutasov duality to this background vector multiplet.

\subsection{Coupling to Background Vector Multiplet}
\label{bgvector}

Let $\hat{\cV}$ be a background vector superfield  for the flavor $U(1)$ symmetry of the theory $\cT_{k,N}$. The action of the electric theory $\cT_{k_e,N_e}$ coupled to $\hat{\cV}$ is given by
\begin{align}
S_e(\Phi,\cV;\hat{\cV}) &\equiv S_{k_e,N_e}(\Phi,\cV; \hat{\cV}) \ec \label{Se}\\
S_{k,N}(\Phi,\cV; \hat{\cV}) &\equiv \frac{ik}{4\pi} S_{\rm CS}(\cV) + S_{\mathbf{N}}(\Phi,\cV+\hat{\cV}) \ec \label{Skn}
\end{align}
where $S_{\rm CS}$ and $S_{\mathbf{N}}$ were defined in \eqref{Scs} and \eqref{SPhi}. Below, we will always use a hat to denote a background field. 

In general, if we demand that our theory be gauge invariant in the background vector fields that source global symmetry currents, then we must add certain Chern-Simons terms in these background fields. This is due to the parity anomaly \cite{Niemi:1983rq,Redlich:1983kn,Redlich:1983dv}. 
If we also insist on supersymmetry, then we need to add supersymmetric Chern-Simons terms in the background vector superfields. The same terms are necessary for the validity of the supersymmetric duality \cite{Benini:2011mf}. Indeed, these global Chern-Simons terms generate contact-terms in the correlation functions of currents; these terms must be added to the dual theories appropriately such that the contact terms match under the duality. In our case, we can account for the contact terms in the 2-point function of the flavor current multiplet by shifting the action of the magnetic theory by
\begin{align}
\delta S_m &= \frac{i k_{FF}}{2\pi} S_{\rm CS}(\hat{\cV}) = \frac{ik_{FF}}{4\pi} \int d^3x \left[ \epsilon_{\mu\nu\rho} \hat{A}^{\mu}\partial^{\nu} \hat{A}^{\rho} + \hat{\bar{\lambda}}\hat{\lambda} + 2i \hat{\sigma}\hat{D}\right]  \ec \label{FFct}\\
 k_{FF} &\equiv -\frac{\sgn(k_m)}{2}\left(|k_m|+\frac{1}{2}\right) \ed \label{kFF}
\end{align}
Here, we are only interested in the effect of these terms on the duality, so for convenience we moved the total contribution to the magnetic theory. The subscript $FF$ denotes the fact that these terms affect the Flavor-Flavor 2-point function.
The global Chern-Simons terms are determined on both sides of the duality by the parity anomaly, up to an integer shift.
The integer part can be determined (for example) by comparing the $S^3$ partition function on both sides of the duality \cite{Closset:2012vp}.
Notice that the global Chern-Simons terms include a term proportional to $\hat{\sigma} \hat{D}$, which shifts the value of the correlator $\langle \Ob \Of \rangle_{\cT}$.
Therefore, another way to determine $k_{FF}$ is to demand that the correlator $\langle \Ob \Of \rangle_{\cT}$ (a pure contact term) maps correctly under the duality.\footnote{
The value of $k_{FF}$ can also be obtained from a perturbative calculation of $\langle\Ob\Of\rangle_{\cT_{k_e,N_e}}$, which gives $\frac{k_e \sin^2 \left( \pi\lambda_e/2 \right)}{4\pi}$ in the planar limit.
Indeed, this result is invariant under the duality only if we shift the magnetic theory action by $\int d^3x\, \hat{\sigma}\hat{D}$ with the coefficient given in  \eqref{FFct}, \eqref{kFF}.
}

In the $\cT$ theory, the Chern-Simons term \eqref{FFct} only serves to ensure that contact terms in certain 2-point functions agree under the duality.
However, our next move will be to define a new theory, $\widetilde{\cT}$, by making the field $\hat{D}$ dynamical.
In this theory the term $\hat{D} \hat{\sigma}$ becomes dynamical and affects correlators at separated points.
Therefore, it is important to correctly add the Chern-Simons term to the $\cT$ theory before proceeding.

Taking the global Chern-Simons term \eqref{FFct} into account, the action of the magnetic theory is given by
\begin{align}
S_m(\Phi,\cV; \hat{\cV}) &= S_{k_m,N_m}(\Phi,\cV;-\hat{\cV}) + \frac{i \kappa_{FF} }{2\pi} S_{\rm CS}(\hat{\cV}) \ed \label{Sm}
\end{align}
Giveon-Kutasov duality then implies the following identity for the partition function $Z_{k,N}[\hat{\cV}]$ of the theory $\cT_{k,N}$:
\begin{align}
Z_{k,N}[\hat{\cV}] &= Z_{-k,|k|-N+1/2}[-\hat{\cV}] \times \exp\left[-\frac{ i \kappa_{FF} }{2\pi} S_{\rm CS}(\hat{\cV})\right] \ec  \label{TknMap}\\
Z_{k,N}[\hat{\cV}] &\equiv \int D\Phi D\cV \, e^{-S_{k,N}(\Phi,\cV;\hat{\cV})} \ed
\end{align}
The identity \eqref{TknMap} exhibits the equivalence $\cT_{k,N}\simeq\cT_{-k,|k|-N+1/2}$ at level of correlators of the current multiplet, obtained by taking derivatives with respect to $\hat{\cV}$.

\subsection{General Bosonization Argument}
\label{general}

The general derivation of 3d bosonization from Giveon-Kutasov duality proceeds as follows. Consider the supersymmetric theory $\cT_{k,N}$ coupled to a background vector multiplet $\hat{\cV}$ for the flavor $U(1)$ symmetry. The $\hat{D}$ (top) component of $\hat{\cV}$ acts as a source for $\Ob=\bar{\vphi}\vphi$ (see \eqref{SPhi}). Let us introduce a term $- \int d^3x \frac{1}{4g} \hat{D}^2$ in the action, and make $\hat{D}$ dynamical. This is equivalent to adding a double-trace $\Ob^2$ deformation, via the Hubbard-Stratonovich trick.
To emphasize that $\hat{D}$ is now dynamical we change our notation for it by removing its hat: $\hat{D}\to D$. We also introduce a source $\hat{B}_0$ for the new dynamical field $D$. We preform this deformation on both sides of the duality by multiplying both sides of equation \eqref{TknMap} by 
\begin{align}
  \exp\left[-\int \! d^3x 
  \left( - \frac{1}{4g}D^2 - k_e \hat{B}_0 D\right)\right] = 
  \exp\left[-\int \! d^3x 
  \left(- \frac{1}{4g}D^2 + k_m \hat{B}_0 D\right)\right] \ec
\end{align}
and path-integrating over $D$. We then flow to the IR fixed points. The resulting non-supersymmetric CFTs in the IR will be denoted by $\widetilde{T}_{k,N}$. In the $\widetilde{T}_{k,N}$ theories $\frac{1}{4g}D^2$ is irrelevant and can be dropped (at least at large $N$).

Notice that $D$ now appears linearly in both the electric and magnetic actions. In the electric theory \eqref{Se} the path integral over $D$ leads to the constraint, $\Ob = k_e\hat{B}_0$. On the other hand, in the magnetic theory \eqref{Sm} the constraint we obtain is $\Ob = k_m\hat{B}_0 - \frac{k_{FF}}{2\pi}\hat{\sigma}$, due to the contact term \eqref{FFct}; this difference will be crucial to the derivation of the correct duality map. Plugging the constraints back into the actions of the electric and magnetic theories, we find\footnote{We have made the change of variables $D\to-D$ in the magnetic theory.}
\begin{align}
\widetilde{S}_e &= \widetilde{S}_{k_e,N_e} + \int \! d^3x \left[ \left(4\pi\hat{B}_0+\hat{\sigma}\right)\Of - k_e\hat{B}_0 D + \cdots \right] \ec \label{tSe} \\
\widetilde{S}_m &= \widetilde{S}_{k_m,N_m} + \int \! d^3x \left[ \left(4\pi\hat{B}_0 - \left(1+\frac{2k_{FF}}{k_m}\right)\hat{\sigma}\right)\Of - \left(k_m\hat{B}_0 -\frac{k_{FF}}{2\pi}\hat{\sigma}\right)D + \cdots \right] \ec \label{tSm}
\end{align}
where the action $\widetilde{S}_{k,N}$ is given by
\begin{align}
\widetilde{S}_{k,N} = \frac{ik}{4\pi} S_{\rm CS}(A) + S_b(\vphi) + S_f(\psi) + \int \! d^3x \left[ D\Ob - \frac{2\pi i}{k} (\bar{\psi}\vphi)(\bar{\vphi}\psi) \right] \ec \label{tTkN}
\end{align}
and $S_{\rm CS}$ , $S_b$ and $S_f$ were defined in \eqref{SCS}, \eqref{Sbb} and \eqref{Sff}, respectively. The omitted terms in \eqref{tSe} and \eqref{tSm} contain additional couplings of sources to the operators $(\bar{\vphi}\psi)$ and $\tilde{J}_{\mu}$ (defined in \eqref{JsSUSYbf}), as well as terms that depend only on the background fields. Those terms will not be important for us. The $(\bar{\psi}\vphi)(\bar{\vphi}\psi)$ interaction is expected to be exactly marginal in the planar limit. In addition, it does not affect planar correlators of bosonic single-trace operators, and we can therefore ignore it for our purposes.

Moving on, we define the partition function of the $\widetilde{\cT}_{k,N}$ theory
\begin{align}
\widetilde{Z}_{k,N}[\hat{B}_0^b, \hat{B}_0^f] \equiv \int \cD \varphi \, \cD \psi \, \cD A \, \cD D \,  e^{-\widetilde{S}_{k,N} -\int \! d^3x \left( \hat{B}_0^b D + \hat{B}_0^f\Of \right)}\ed
\end{align}
At large $N$, from \eqref{kFF} we see that $k_{FF} \to - k_m/2$, and the $\cN=2$ duality \eqref{TknMap} implies the identity
\begin{align}
\widetilde{Z}_{k,N}\left[-k\hat{B}_0, 4\pi\left(\hat{B}_0+\frac{1}{4\pi}\hat{\sigma}\right)\right] = \widetilde{Z}_{-k,|k|-N}\left[k\left(\hat{B}_0+\frac{1}{4\pi}\hat{\sigma}\right), 4\pi\hat{B}_0\right] \ed \label{tTknMap}
\end{align}
The relation \eqref{tTknMap} should be understood as an identity of correlators of $D$ and $\Of$ at separated points, obtained by taking derivatives w.r.t. $\hat{B}_0$ and $\hat{\sigma}$. In particular, we find that under $\widetilde{\cT}_{k,N}\to\widetilde{\cT}_{-k,|k|-N}$, the operators $\Of$ and $D$ get mapped to each other according to the prescription
\begin{align}
D \to - \frac{4\pi}{k} \Of \ecq \Of \to \frac{k}{4\pi} D \ed \label{DOfmap}
\end{align}
This agrees with the mapping \eqref{Otmap} that was derived using perturbation theory in the multi-trace couplings of the $\cN=2$ theory (note that $D$ is equal to $\tOb/N$).

In the planar limit the above duality can be directly related to the 3d bosonization duality between the boson theory $\cBcrit_{k,N}$ and the  fermion theory $\cF_{-k,|k|-N}$, whose actions were given in \eqref{Scb} and \eqref{Sf} . Indeed, it is not hard to verify that at the level of planar correlators of the operators $D$ and $\Of$, the theory $\widetilde{\cT}_{k,N}$ factorizes into a decoupled product of the $\cBcrit_{k,N}$ and $\cF_{k,N}$ CFTs. In particular,\footnote{One way to obtain \eqref{TB} and \eqref{TF} is to integrate out the gauge field $A_{\mu}$ in the theory $\widetilde{\cT}_{k,N}$ in light-cone gauge \cite{Jain:2012qi}. The  interactions between the fermion $\psi$ and the scalars $\vphi$ and $D$ in the resulting non-local action have no effect on planar correlation functions of $D$ and $\Of$.}
\begin{align}
  \langle D(x_1)\cdots D(x_n) \rangle_{\widetilde{\cT}_{k,N}}^{\rm planar} &= \langle D(x_1)\cdots D(x_n) \rangle_{\cBcrit_{k,N}}^{\rm planar} \ec \label{TB}\\
  \langle \Of(x_1)\cdots \Of(x_n) \rangle_{\widetilde{\cT}_{k,N}}^{\rm planar} &= \langle \Of(x_1)\cdots \Of(x_n) \rangle_{\cF_{k,N}}^{\rm planar} \ed \label{TF}
\end{align}
Another way to reach this conclusion is to turn on a mass for the fermion. In the planar limit the scalar propagator does not receive corrections from the fermion, so the scalar remains massless and we flow to the critical bosonic theory in the IR. Under the duality \eqref{DOfmap}, the deformation maps to a relevant deformation involving the bosonic $D$ which does not correct the fermion propagator. In the IR the scalar decouples, and we flow to the fermionic theory.

We conclude that under $\cBcrit_{k,N}\to\cF_{-k,|k|-N}$, the operators $D$ and $\Of$ also map to each other according to \eqref{DOfmap}. In fact, one can verify that the $k/4\pi$ factor in \eqref{DOfmap} agrees with known results for 2-point and 3-point functions \cite{Aharony:2012nh,GurAri:2012is}. Note that correctly accounting for the $\cN=2$ duality map of the contact term in the 2-point function of the flavor $U(1)$ current was crucial in deriving this factor. We view the above arguments as a proof that \eqref{DOfmap} must hold in any $n$-point function of $D$ and $\Of$ in the theories $\cBcrit_{k,N}$ and $\cF_{k,N}$, given that Giveon-Kutasov duality of the theory $\cT_{k,N}$ is correct.

\subsection{Including Currents}

The above arguments can be easily generalized to include correlators of the other single-trace operators $\Jb_s$ and $\Jf_s$, given in \eqref{Jb} and \eqref{Jf}. We simply couple these operators to sources in the $\cN=2$ action and follow the same derivation leading to \eqref{tTknMap}. In this case the duality map in the $\widetilde{T}_{k,N}$ theory is the same as the one of the $\cN=2$ theory, which was given in \eqref{TknOpMap}. As before, from the point of view of planar correlators of $D$, $\Jb_s$, $\Of$ and $\Jf_s$, the $\widetilde{T}_{k,N}$ CFT factorizes into a decoupled product of the $\cBcrit_{k,N}$ and $\cF_{k,N}$ theories. Therefore under $\cBcrit_{k,N}\to\cF_{-k,|k|-N}$, we must have that $\Jb_s \to (-)^s \Jf_s$ in agreement with \eqref{Otmap}.

There is an important loophole in the above argument that we must address. It is possible that for the $\cN=2$ duality to be valid in the presence of sources for the currents, one must shift the action of the magnetic theory by a local functional of those sources and of $\hat{D}$. In the presence of such terms the constraint imposed by integrating over $\hat{D}$ would be modified, and our conclusions could be invalidated. 
Indeed, the flavor-flavor contact term \eqref{FFct}, which includes a term proportional to $\hat{D} \hat{\sigma}$, was crucial in deriving the duality map \eqref{TknOpMap}.
We will now show that it is not possible to write another local functional of the sources that would end up contributing to correlators in the bosonic and fermionic theories at separated points.

To see this, let us denote the sources of $\Jb_s$ and $\Jf_s$ by $\hat{B}_s^b$ and $\hat{B}^f_s$.
We take these tensors to be symmetric and traceless. 
The local functionals we consider are of the form $S_{\rm c.t.}(\hat{D},\hat{\sigma},\hat{B}_s^b,\hat{B}_s^f,\dots)$, where $(\dots)$ denotes the fundamental fields, and where all terms are at least quadratic in the sources.
First note that we only have to consider functionals that are at most linear in $\hat{\sigma}$ and $\hat{B}^{b,f}_s$, but can otherwise have any positive power of $\hat{D}$. This is because non-linear terms in $\hat{\sigma}$ and $\hat{B}^{b,f}_s$ would only affect contact terms in the transformed theories (in which $\hat{D}$ is dynamical).
We will show that there are no such terms that include a factor of $\hat{B}^{b,f}_s$; similar considerations rule out terms that involve only $\hat{D}$, or both $\hat{D}$ and $\hat{\sigma}$ (except for the term $\hat{D}\hat{\sigma}$).
The most general local functional we can write down, which satisfies the above requirements, is
\begin{align}
  \int \! d^3x \, \hat{D}^n \cO^{\mu_1 \dots \mu_s} \hat{B}^{b,f}_{\mu_1 \dots \mu_s}
   \ed
\end{align}
Here $n>0$, and $\cO$ is an operator of dimension $\Delta$ and spin $s$. The operator $\cO$ may be any product of a local operator with a differential operator whose derivatives act on $\hat{B}^{b,f}$, but its particular form will not be important. 
One can now easily check that the twist of $\cO$ is $\Delta - s = 1-2n < 0$.
In order to have negative twist, $\cO$ must include factors of $\delta_{\mu\nu}$ or $\epsilon_{\mu\nu\rho}$, but then the counter-term vanishes by assumption (the sources are assumed to be symmetric and traceless).
This concludes the proof.

\section{Theories with One Boson and One Fermion}
\label{BosFer}

In this section we will derive the duality map of various supersymmetry breaking deformations of the $\cN=2$ theory. In particular, we reproduce the duality map presented in \cite{Jain:2013gza} for the theory with both a scalar and a fermion, and also extend their results to other deformations.

\subsection{$\cN=2\to\cN=1$}

Let us start by breaking $\cN=2$ supersymmetry only partially, such that we obtain an $\cN=1$ duality. To do that we first rewrite the action \eqref{Skn} of the $\cN=2$ theory $\cT_{k,N}$ in $\cN=1$ language.\footnote{More details on the $\cN=1$ decomposition of $\cN=2$ Chern-Simons matter actions can be found in \cite{Ivanov:1991fn}.} The $\cN=2$ chiral superfield $\Phi(x,\theta,\bar{\theta})$ can be written in terms of an $\cN=1$ complex scalar superfield $\phi(x,\theta)$. The $\cN=2$ vector multiplet $\cV(x,\theta,\bar{\theta})$ decomposes into an $\cN=1$ vector multiplet $\Gamma_{\alpha}(x,\theta)$ plus a real scalar multiplet $B(x,\theta)$. Similarly, we will denote the $\cN=1$ components of the background vector multiplet $\hat{\cV}$, by $\hat{\Gamma}_{\alpha}$ and $\hat{B}$. The superfield $B$ is auxiliary and can be integrated out. After $B$ has been eliminated the action of $\cT_{k,N}$, defined in \eqref{Skn}, can be written in terms of the remaining $\cN=1$ variables as
\begin{align}
S_{k,N}(\phi,\Gamma_{\alpha};\hat{\Gamma}_{\alpha},\hat{B}) &= \frac{ik}{4\pi} S_{\rm CS}^{\cN=1}(\Gamma_{\alpha}) + \int \! d^3x \, d^2\theta\left[\cD^{\alpha}\bar{\phi}\cdot\cD_{\alpha}\phi - \frac{2\pi i}{k}(\bar{\phi}\phi)^2\right] \notag\\
& + \int \! d^3x d^2\theta \left[\hat{\Gamma}^{\alpha}\left( i\bar{\phi}\cD_{\alpha}\phi - i\left(\cD_{\alpha} \bar{\phi}\right)\phi + \hat{\Gamma}_{\alpha}\bar{\phi}\phi\right) + \hat{B}\bar{\phi}\phi \right] \ec \\
S_{\rm CS}^{\cN=1}(\Gamma_{\alpha}) &\equiv -\frac{1}{2}\int \! d^3x\, d^2\theta \tr\left[ D^{\alpha}\Gamma^{\beta}D_{\alpha}\Gamma_{\beta} - \frac{2i}{3}D^{\alpha}\Gamma^{\beta}\{\Gamma_{\alpha},\Gamma_{\beta}\} - \frac{1}{6}\{\Gamma^{\alpha},\Gamma^{\beta}\}\{\Gamma_{\alpha},\Gamma_{\beta}\} \right] \ec
\end{align}
where $\cD_{\alpha}\phi\equiv D_{\alpha}\phi -i\Gamma_{\alpha}\phi$ and  $\cD_{\alpha}\bar{\phi}\equiv D^{\alpha}\bar{\phi} + i \bar{\phi}\,\Gamma^{\alpha}$.
Moreover, in terms of $\cN=1$ variables the abelian Chern-Simons term $S_{\rm CS}(\hat{\cV})$ that appears on the RHS of the identity \eqref{TknMap} is given by
\begin{align}
S_{\rm CS}(\hat{\cV}) &= \frac{i}{8}\int \! d^3x d^2\theta\left[ D^{\alpha}\hat{\Gamma}^{\beta}D_{\alpha}\hat{\Gamma}_{\beta} - 2 \hat{B}^2\right]\ec \label{SmN1}
\end{align}

Let us now multiply both sides of the identity \eqref{TknMap} by $e^{-\delta S}$ with $\delta S \equiv \frac{k\mu}{2\pi(w-1)}\hat{B} + \frac{k}{8\pi i(w-1)}\hat{B}^2$, and then path-integrate boths sides over $\hat{B}$. The choice of parameters in $\delta S$ is such that $\mu$ and $w$ coincide with the definitions of \cite{Jain:2013gza}. After $\hat{B}$ has been eliminated by using its equations of motion, we are left with an identity that exhibits the self-duality of an $\cN=1$ $U(N)$ Chern-Simons theory coupled to a fundamental scalar superfield $\phi$ with an arbitrary renormalizable superpotential. The action of this $\cN=1$ theory is given by
\begin{align}
S^{\cN=1}_{k,N}(\phi,\Gamma_{\alpha}) = \frac{ik}{4\pi} S_{\rm CS}^{\cN=1}(\Gamma_{\alpha}) + \int \! d^3x\, d^2\theta \left[ \cD^{\alpha}\bar{\phi}\cdot\cD_{\alpha}\phi - 2 i \mu(\bar{\phi}\phi) - \frac{2\pi i}{k} w (\bar{\phi}\phi)^2 \right]\ed \label{SknN1}
\end{align}
The duality map of the parameters in \eqref{SknN1} is found to be\footnote{One can also read off the duality map of the $U(1)$ current that couples to background $\cN=1$ gauge field $\hat{\Gamma}_{\alpha}$.}
\begin{align}
N \to |k|-N \ecq k \to -k \ecq \mu \to -\frac{2\mu}{1+w} \ecq w \to \frac{3-w}{1+w} \ec
\end{align}
where we set the coefficient $\kappa_{FF}$ of the contact-term action in \eqref{TknMap} to its large $N$ value: $\kappa_{FF} \to -k_m/2$. This is precisely the duality map for $\mu$ and $w$ that was found in \cite{Jain:2013gza}. 

\subsection{$\cN=2\to\cN=0$}

The same reasoning that carried us so far can be used to obtain the bosonization duality map for the most general renormalizable Chern-Simons vector model with one scalar and one fermion. We multiply the identity \eqref{TknMap} by $e^{-\delta S}$, where $\delta S$ is now the most general renormalizable functional of the auxiliary fields $\hat{\sigma}$, $\hat{\lambda}$, $\hat{\bar{\lambda}}$ and $\hat{D}$ in the background vector multiplet. In particular, $\delta S$ is given by
\begin{align}
\delta S = \int \! d^3x \, \left[ \alpha_1 \hat{\sigma} + \alpha_2 \hat{\sigma}^2 + \alpha_3 \hat{\sigma}^3 + \beta_1 \hat{D} + \beta_2 \hat{\sigma} \hat{D} + \gamma_1 \hat{\bar{\lambda}}\hat{\lambda} + \gamma_2\left(\hat{\lambda}^2 + \hat{\bar{\lambda}}^2\right) \right] \ed \label{dSN0}
\end{align}
The full action of the electric theory, after integrating out the auxiliary fields of the dynamical vector multiplet $\cV$, can be written as
\begin{align}
  S^{\cT}_{k,N}(A_\mu,\varphi,\psi) + \delta S +
  \int \! d^3x \, \left[ 
  \hat{D} \Ob + \hat{\sigma} \left( \Of + \frac{4\pi}{k} \Ob^2 \right) +
  \hat{\sigma}^2 \Ob + i\hat{\lambda} \chi + i\hat{\bar{\lambda}} \bar{\chi}
  \right] \ed
  \label{STelec}
\end{align}
Here, we set the background gauge field $\hat{A}$ to zero, and defined $\chi \equiv \bar{\vphi}\psi$ and $\bar{\chi}\equiv \bar{\psi}\vphi$.
The action of the magnetic theory is
\begin{align}
  &S^{\cT}_{-k,|k|-N}(A_\mu,\varphi,\psi) + \delta S +
  \int \! d^3x \, \left[ 
  - \hat{D} \Ob - \hat{\sigma} \left( \Of - \frac{4\pi}{k} \Ob^2 \right) +
  \hat{\sigma}^2 \Ob - i\hat{\lambda} \chi - i\hat{\bar{\lambda}} \bar{\chi}
  \right]
  \cr
  &+ \frac{i \kappa_{FF}}{4\pi} \int \! d^3x \, \left[ 
  \hat{\bar{\lambda}} \hat{\lambda} + 2 i \hat{\sigma} \hat{D}
  \right] \ed
  \label{STmag}
\end{align}
We now integrate over the background auxiliary fields on both sides of the identity \eqref{TknMap}, where the full actions on both sides are given by \eqref{STelec},\eqref{STmag}. To match our conventions with those of \cite{Jain:2013gza} we introduce the parameters $m_b$, $m_f$, $b_4$, $x_4$, $x_6$, $y_4'$ and $y_4''$, and identify these with the parameters in $\delta S$ according to
\begin{align}
\alpha_1 &= \frac{k  \left[m_f \left( 8 b_4 (1-x_4)+m_f(1-4  x_4+3 x_6)\right)+4 m_b^2 (x_4-1)^2\right]}{16 \pi  (x_4-1)^3} \ec \\
\alpha_2 &= \frac{k \left[4 b_4 (x_4-1)+m_f \left(4 x_4^2-3 x_6-1\right)\right]}{16 \pi  (x_4-1)^3} \ec \\
\alpha_3 &= \frac{k\left[ x_6 + 4 x_4 (1-x_4) -1 \right]}{16\pi(x_4-1)^3} \ec \\
\beta_1 &= \frac{k \,m_f}{4\pi(x_4-1)} \ec \\
\beta_2 &= -\frac{k}{4\pi(x_4-1)} \ec \\
\gamma_1 &= -\frac{ i k}{2\pi} \frac{1+y_4'}{ (1+y_4')^2-4y_4''^2} \ec \\
\gamma_2 &= \frac{i k}{2\pi} \frac{y_4''}{(1+y_4')^2 - 4 y_4''^2} \ed
\end{align}

After the auxiliary fields $\hat{\sigma}$, $\hat{D}$, $\hat{\lambda}$ and $\hat{\bar{\lambda}}$ have been integrated out in the electric theory, we are left with the most general $U(N)$ Chern-Simons theory coupled to a fundamental scalar $\vphi$ and fermion $\psi$, with the matter potential\footnote{We discard a constant shift in the potential $V(\vphi,\psi)$ that is independent of the fields.}
\begin{align}
V(\vphi,\psi) &= m_f O_f + m_b^2 O_b + \frac{4\pi b_4}{k} O_b^2 + \frac{4\pi^2x_6}{k^2} O_b^3 + \frac{4\pi x_4}{k} O_b O_f + \frac{2\pi i y_4'}{k} \bar{\chi}\chi + \frac{2\pi i y_4''}{k}\left(\chi^2+\bar{\chi}^2\right) \ed
\end{align}
Repeating this in the magnetic theory, we find that the self-duality map is
\begin{gather}
k \to -k \ecq x_4 \to \frac{1}{x_4} \ecq x_6 \to 1 + \frac{1-x_6}{x_4^3} \ecq m_f \to -\frac{m_f}{x_4}  \ec \notag \\
m_b^2 \to -\frac{1}{x_4} m_b^2 + \frac{3}{4}\frac{1-x_6}{x_4^3}m_f^2 + \frac{2}{x_4^2} m_f\,b_4  \ecq b_4 \to -\frac{1}{x_4^2}\left(b_4 + \frac{3}{4}\frac{1-x_6}{x_4}m_f\right) \ec \notag\\
y_4' \to 3 + 8\left( \frac{1}{y_4'-2y_4''-3}+\frac{1}{y_4'+2y_4''-3}\right) \ecq y_4'' \to  \frac{16 y_4''}{4y_4''^2 - (y_4'-3)^2} \ed \label{Vtrans}
\end{gather}
The mapping that we found agrees with \cite{Jain:2013gza}. The transformation rules for the couplings $y_4'$ and $y_4''$ are new. Note that the point $x_4=x_6=-y_4'=1$ and $m_b=m_f=b_4=0$ is a fixed point of \eqref{Vtrans}, which corresponds to the $\cN=2$ Giveon-Kutasov duality.

\section{Discussion}
\label{discussion}

Let us summarize our results. We proved that all planar correlators of single-trace operators must map correctly under the 3d bosonization map given in \eqref{Otmap} if the Giveon-Kutasov duality is correct. In the process we have uncovered signs in the duality transformation that were not noticed previously. Moreover, we gave a new derivation of the transformation \eqref{Vtrans} of the most general renormalizable matter potential $V(\vphi,\psi)$ in the Chern-Simons vector model with both a scalar and a fermion; the transformation rule for some of the couplings in $V(\vphi,\psi)$ was not known previously. The main advantage of our approach is that it is simple, and does not rely on making complicated computations. 
 
We exhibited the relation between the $\cN=2$ theory and the non-supersymmetric bosonic and fermionic models in two different ways. In Section \ref{cpt} we showed that planar correlators of the $\cN=2$ theory can be expressed algebraically in terms of correlators of the non-supersymmetric theories.
On the other hand, in Section \ref{dualityProof} we have seen that these theories are related by a double-trace flow, followed by a mass deformation to decouple either the boson or the fermion.
At large $N$, these two approaches are related. For example, the critical $O(N)$ model is related to the free $O(N)$ model by a double-trace flow, and the planar correlators of the critical $O(N)$ model are algebraically related to those of the free model. These relations can be seen by re-summing the perturbative series in the double-trace interaction, similarly to our approach in Section \ref{cpt}. Alternatively, by re-writing the double-trace deformation using the Hubbard-Stratonovich trick, the correlators of the two theories are seen to be simply related by a Legendre transform \cite{Klebanov:1999tb,Witten:2001ua,Gubser:2002vv}; this is similar in spirit to our approach in Sections \ref{dualityProof} and \ref{BosFer}. 

The relations between the $\cN=2$ and non-supersymmetric theories can be used to derive the duality of the latter from that of the former. In order to apply this strategy, we derived the $\cN=2$ duality map of our supersymmetry-breaking double-trace deformation. This map was shown to be related to the known transformation of the flavor $U(1)$ current multiplet, via the Hubbard-Stratonovich trick. A crucial ingredient in the derivation was that we had to extend the $\cN=2$ duality such that it held also for the contact term in the 2-point function of the $U(1)$ current multiplet. 

The manipulations used in Sections \ref{dualityProof} and \ref{BosFer} involved deforming the actions of two dual theories by background fields, and then path-integrating over them. These manipulations are rather formal, and one may ask whether they are still valid after renormalization. Here we rely on the fact that our theories are essentially finite in the planar limit. Indeed, in this limit the R symmetry of the $\cN=2$ theory is not renormalized, and its supersymmetry-breaking multi-trace deformations do not lead to logarithmic divergences. Our path-integral manipulations are therefore completely well defined in the planar limit. 

There are several goals one might hope to achieve through a better understanding of the relation between the $\cN=2$ and non-supersymmetric dualities, which we leave to future work. Most importantly, this understanding could lead to evidence for non-supersymmetric bosonization at finite $N$. The main obstacle to extending our arguments (or those of \cite{Jain:2013gza}) to finite $N$, is that this would require taking renormalization effects into account. Moreover, recall that our argument relied on the decoupling of the Wilson-Fisher scalar and the fermion in the CFT $\widetilde{\cT}_{k,N}$, which arises by flowing to the IR from the $\cN=2$ theory $\cT_{k,N}$. 
At finite $N$ this flow might require a fine-tuning of the classically marginal interactions $(\bar{\psi}\vphi)(\bar{\vphi}\psi)$ and $\left((\bar{\psi}\vphi)^2 + \mathrm{c.c.}\right)$. In the planar limit this subtlety was avoided because these deformations are exactly marginal. It would be interesting to check whether they become relevant or irrelevant at finite $N$.
Even if we could understand the $\cT_{k,N}\to\widetilde{\cT}_{k,N}$ flow at finite $N$, beyond the planar limit it is no longer true that the scalar and fermion are decoupled in $\widetilde{\cT}_{k,N}$. 
We would then need to show that one can flow from $\widetilde{\cT}_{k,N}$ to the bosonic and fermionic models. 

There are other future research directions that are interesting even if we restrict ourselves to the large $N$ limit. Supersymmetry makes it possible to compute many interesting observables, such as partition functions on curved manifolds and correlation functions of Wilson loops. These quantities are currently not known in the non-supersymmetric theories even in the planar limit. It would be interesting if we could use the supersymmetric results to learn about these observables in the non-supersymmetric theories. In adition, there is a large class of dualities in $\cN=2$ Chern-Simons matter theories. It is possible that the simple arguments given in this paper could be extended to find new examples of non-supersymmetric dualities.

There is an additional open question that is related to our study of $n$-point functions in Chern-Simons vector models. These theories are conjectured to be holographically dual to Vasiliev theories of higher-spin gravity in $AdS_4$ \cite{Sezgin:2001zs,Klebanov:2002ja,Giombi:2012ms}, which have an infinite tower of parity violating couplings. One of those bulk couplings was matched with the `t~Hooft coupling on the CFT side \cite{Chang:2012kt}, while the interpretation of the other couplings is unknown. The structure of Vasiliev's equations suggests that those additional parameters may only affect boundary 5-point functions and higher, which have never been computed. If these couplings do have a physical effect then it leads to a puzzle in the holographic duality, because there are no obvious marginal parameters on the CFT side that could correspond to those parameters. In particular, one would expect that the bosonic and fermionic models, that are dual to one another under bosonization, are holographically dual to bulk theories with generally different values of these parameters. The bosonization duality could then fail at the level of planar 5-point functions and higher. Since our results give evidence that all planar $n$-point functions agree under bosonization, they also give indirect evidence that those bulk couplings are not physical. It would be interesting to better understand this issue, for example by counting solutions to the conformal bootstrap, as was done in \cite{Heemskerk:2009pn}, but for theories with slightly broken higher-spin symmetry.

\subsection*{Acknowledgments}
\label{s:acks}

We thank Ofer Aharony, Cyril Closset, Ethan Dyer, Zohar Komargodski, Juan Maldacena, and Itamar Yaakov  for useful discussions.  
The work of RY was supported in part by the US NSF under Grant No. PHY-1314198.
The work of GGA was supported by a grant from the John Templeton Foundation. The opinions expressed in this publication are those of the authors and do not necessarily reflect the views of the John Templeton Foundation.

\appendix

\section{Conventions}
\label{conv}

In this appendix we collect some details on our conventions regarding $\cN=2$ supersymmetry in 3d. We work in 3d Euclidean space with flat metric  $\delta_{\mu\nu}={\rm diag}(1,1,1)$, where $\mu,\nu=1,2,3$. The Dirac matrices are defined to be the usual Pauli matrices, $(\gamma^{\mu})_{\alpha}{}^{\beta}=\sigma^{\mu}$, where $\alpha,\beta=1,2$.  Spinor indices are raised and lowered from the left with the antisymmetric tensors $\varepsilon_{\alpha\beta}$ and $\varepsilon^{\alpha\beta}$, where $\varepsilon_{12}=-\varepsilon^{12} = -1$. When indices are suppressed their contraction is defined using the North-West to South-East convention,
\begin{align}
\psi\xi \equiv \psi^{\alpha}\xi_{\beta} = \xi\psi \ecq
\psi \gamma^\mu \xi \equiv
\psi^\alpha (\gamma^\mu)_\alpha^{\;\;\beta} \xi_\beta =
- \xi \gamma^\mu \psi \ed
\end{align}

The spinors $\psi$ and $\bar{\psi}$ are independent in Euclidean space, whereas in Minkowski space they would be hermitian conjugates. In particular, the Grassmann coordinate on $\cN=2$ superspace are given by two independent complex spinors $\theta^{\alpha}$ and $\bar{\theta}^{\alpha}$. The supersymmetric covariant derivatives are defined by
\begin{align}
D_{\alpha} = \frac{\partial}{\partial\theta^{\alpha}} -i(\gamma^{\mu}\bar{\theta})_{\alpha}\partial_{\mu} \ecq \bar{D}_{\alpha} = -\frac{\partial}{\partial\bar{\theta}^{\alpha}} +i(\gamma^{\mu}\theta)_{\alpha}\partial_{\mu} \ec
\end{align}
and satisfy the algebra
\begin{align}
\{D_{\alpha},D_{\beta}\} = \{\bar{D}_{\alpha},\bar{D}_{\beta}\} = 0 \ecq \{D_{\alpha},\bar{D}_{\beta}\} = - 2 i \partial_{\alpha\beta} \ed
\end{align}

In order to construct supersymmetric actions we use the following conventions for superfields. Chiral superfields $\Phi(x,\theta,\bar{\theta})$ are defined by the constraint $\bar{D}_{\alpha}\Phi = 0$ and can be written in components as

\begin{align}
	\Phi &= \vphi(x) + \sqrt{2}\theta\psi(x) + i \bar{\theta}\gamma^{\mu}\theta\partial_{\mu}\vphi(x) - \frac{i}{\sqrt{2}}\theta^2\bar{\theta}\gamma^{\mu}\partial_{\mu}\psi(x) - \frac{1}{4}\theta^2\bar{\theta}^2\Box\!\vphi+\theta^2F(x) \ed \label{chiral}
\end{align}
Similarly, anti-chiral superfields $\bar{\Phi}$ satisfy $D_{\alpha}\bar{\Phi}=0$, and are given in components by
\begin{align}
	\bar{\Phi} 	&= \bar{\vphi}(x) - \sqrt{2}\bar{\theta}\bar{\psi}(x) - i \bar{\theta}\gamma^{\mu}\theta\partial_{\mu}\bar{\vphi}(x) - \frac{i}{\sqrt{2}}\bar{\theta}^2\partial_{\mu}\bar{\psi}(x)\gamma^{\mu}\theta - 	\frac{1}{4}\theta^2\bar{\theta}^2\Box\!\bar{\vphi} - \bar{\theta}^2\bar{F}(x) \ed \label{antichiral}
\end{align}
A vector multiplet is described by a real superfield $\cV$, $\cV^{\dagger}=\cV$, whose components in Wess-Zumino gauge are
\begin{align}
	\cV = \theta\gamma^{\mu}\bar{\theta}A_{\mu} - i \theta\bar{\theta}\sigma - \frac{i}{\sqrt{2}}\theta^2\bar{\theta}\bar{\lambda} + \frac{i}{\sqrt{2}}\bar{\theta}^2\theta\lambda - \frac{1}{2}\theta^2\bar{\theta}^2D \ed \label{vector}
\end{align}
Integration over superspace is defined by
\begin{align}
\int d^2\theta \, \theta^2 = \int d^2 \bar{\theta} \, \bar{\theta}^2 = \int d^4\theta \,\theta^2 \bar{\theta}^2 = 1 \ed
\end{align}

\section{Proof that $\langle JO \rangle$ Correlators Vanish Exactly}
\label{2ptapp}

In this section we prove that correlators of the form $\langle JO \rangle$, where $J$ is a current and $O$ is a scalar, vanish exactly in the planar limit.
The proof holds when all the finite counter-terms that affect contact terms of 2-point functions are set to zero (see Section \ref{cptPrel}).

Let $\Jf_s$ be a current with spin $s>0$, and let $O_f$ be the scalar operator in the fermionic theory $\cF_{k,N}$.
The correlator $\langle \Jf_s O_f \rangle$ vanishes at separated points because of conformal symmetry.
We will show that the correlator $\langle \Jf_s O_f \rangle$ vanishes exactly in the planar limit, even at coincident points.
In other words, we will prove that this correlator does not contain contact terms.
This is true both in the fermionic and in the $\cN=2$ Chern-Simons vector models, and a similar result will be shown for correlators of the form $\langle \Jb_s O_b \rangle$ in the regular bosonic and $\cN=2$ theories.
Using perturbation theory, it is then easy to check that $\langle \Jf_s \Ob \rangle$ and $\langle \Jb_s \Of \rangle$ also vanish in the $\cN=2$ theory.

\subsection{Fermionic Case}

In this section we distinguish between correlators that vanish only up to contact terms, and correlators that vanish exactly.
Regarding the operator $\Jf_s$, we assume that it is symmetric, conserved and traceless inside any planar 2-point function of single-trace operators, but only up to contact terms.\footnote{Note that our currents are generally conserved and traceless only up to multi-trace operators, but such operators do not affect the planar 2-point functions we are considering here.
Therefore, for the purposes of this section the currents are conserved and traceless, at least at separated points.}
Consider the momentum-space correlator $\langle \Jf_s(p) O_f \rangle_{\cF}$ in the fermionic theory.
The most general form it can take is
\begin{align}
  \langle \Jf_{\mu_1 \cdots \mu_s}(p) O_f \rangle_{\cF} =
  P_{\mu_1 \cdots \mu_s}(p) \ec
  \label{poly}
\end{align}
where $P_{\mu_1\cdots\mu_s}(p)$ is a polynomial in the momentum $p_\mu$ of dimension $s$, corresponding to contact terms in the $x$-space correlator.  

The corresponding correlator in the supersymmetric $\cN=2$ theory is given by
\begin{align}
  \langle \Jf_{\mu_1 \cdots \mu_s}(p) O_f \rangle_{\cT_{k,N}} =
  \Bigg[ 1 - \frac{4\pi}{k} \langle O_b(p) O_f \rangle_{\cT_{k,N}} \Bigg] \langle \Jf_{\mu_1 \cdots \mu_s}(p) O_f \rangle_{\cF_{k,N}} = 
  c \langle \Jf_{\mu_1 \cdots \mu_s}(p) O_f \rangle_{\cF_{k,N}} \ec
\end{align}
where $c$ is a non-vanishing function of $N,k$.
Here we are using the fact that the 2-point function in the supersymmetric theory can be written in terms of correlators of the non-supersymmetric theories.
This can be seen by using a perturbative expansion in the $(\bar{\varphi} \varphi) (\bar{\psi} \psi)$ vertex of the supersymmetric theory, as discussed in Section \ref{cptPrel}. 
Therefore, in order to show that the correlator vanishes in both theories it is enough to show it for the fermionic theory.

Let us now prove that the polynomial $P(p)$ vanishes.
First, let us show that $P_{\mu_1 \cdots \mu_s}(p)$ is symmetric, conserved and traceless.
To see this, consider the 2-point function of the current in the supersymmetric theory.
We can write it as\footnote{In this section we consider only dynamical contributions to correlators, ignoring contributions due to counter-terms that are composed of background fields.
This statement includes correlators that show up as sub-diagrams in other correlators, such as the correlator $\langle O_b O_b \rangle_{\cT}$ in \eqref{JfJf}.
}
\begin{align}
  \langle \Jf_{\mu_1 \cdots \mu_s}(p) \Jf_{\nu_1 \cdots \nu_s} \rangle_{\cT_{k,N}}
  &=
  \langle \Jf_{\mu_1 \cdots \mu_s}(p) \Jf_{\nu_1 \cdots \nu_s} \rangle_{\cF_{k,N}}
  + 
  \cr &\quad
  \left( \frac{4\pi}{k} \right)^2 
  \langle \Jf_{\mu_1 \cdots \mu_s}(p) O_f \rangle_{\cF_{k,N}} \cdot
  \langle O_b(p) O_b \rangle_{\cT_{k,N}} \cdot
  \langle O_f(p) \Jf_{\nu_1 \cdots \nu_s} \rangle_{\cF_{k,N}} \ed \cr
  \label{JfJf}
\end{align}
Conformal symmetry implies that $\langle O_b(p) O_b \rangle \propto |p|^{-1}$. 
Using also \eqref{poly}, the 2-point function can be written as
\begin{align}
  \langle \Jf_{\mu_1 \cdots \mu_s}(p) \Jf_{\nu_1 \cdots \nu_s} \rangle_{\cT_{k,N}}
  &= 
  \langle \Jf_{\mu_1 \cdots \mu_s}(p) \Jf_{\nu_1 \cdots \nu_s} \rangle_{\cF_{k,N}} +
  \frac{c'}{|p|} P_{\mu_1 \cdots \mu_s}(p) 
  P_{\nu_1 \cdots \nu_s}(p) \ec
\end{align}
where $c'$ is a non-vanishing function of $N,k$.
By assumption, the correlators $\langle \Jf \Jf \rangle$ in both the fermionic and supersymmetric theories are conserved at separated points.
Therefore,
\begin{align}
  \frac{1}{|p|} p^{\mu_1} P_{\mu_1 \cdots \mu_s}(p) P_{\nu_1 \cdots \nu_s}(p)
  = {\rm polynomial~in~}p \ed
\end{align}
This is only possible if $P_{\mu_1\cdots\mu_s}$ is conserved.
A similar argument shows that $P_{\mu_1\cdots\mu_s}$ is symmetric and traceless. 

We conclude that $P_{\mu_1\cdots\mu_s}(p)$ is a conserved, symmetric, and traceless tensor of dimension $s$. It is easy to see that such an object must vanish. Indeed, define $P_s(p;y)\equiv y^{\mu_1}\cdots y^{\mu_s}P_{\mu_1\cdots\mu_s}(p)$, where the $y$ are commuting and null polarizations (i.e., $y\cdot y=0$). Since $P_{\mu_1\cdots\mu_s}(p)$ is symmetric and traceless, it is uniquely determined from $P_s(p;y)$. Moreover, because $P_s$ has dimension $s$ and the $y$ are null, it can only take the form $P_s = c \,(y\cdot p)^s$, for some constant $c$. Finally, by imposing conservation, $p\cdot\partial_y P_s(p;y) = 0$, we conclude that $c=0$. 

\subsection{Bosonic Case}

In this section we prove that $\langle \Jb_s O_b \rangle$ vanishes in both the regular bosonic (without a $(\bar{\varphi} \varphi)^2$ deformation) and supersymmetric $\cN=2$ theories, in the planar limit.
Here $\Jb_s$ is a current with $s>1$, and $O_b = \bar{\varphi} \varphi$ is the bosonic scalar operator.
As in the fermionic case, the correlator of the bosonic theory is proportional to the one in the supersymmetric theory, and therefore it is enough to show that the bosonic correlator vanishes.
The most general form of this correlator in momentum space is
\begin{align}
  \langle \Jb_{\mu_1 \cdots \mu_s}(p) O_b \rangle = 
  Q_{\mu_1 \cdots \mu_s}(p) \ec
\end{align}
where $Q$ is again a polynomial in $p$.
$Q$ has dimension $s-1$, which implies that it must include a factor of $\epsilon$.
If $Q$ is symmetric then no such term can be written down, and therefore $Q$ vanishes.
It is left to show that $Q$ is symmetric, and this can again be shown by considering the 2-point function of the current in the supersymmetric theory.
This concludes the proof.

\end{document}